\documentclass[%
 aip,
 amsmath,amssymb,
 pof,
preprint,%
author-year,%
]{revtex4-1}

\usepackage{natbib}
\usepackage{graphicx}
\usepackage{dcolumn}
\usepackage{amsmath}
\usepackage{bm}

\usepackage[utf8]{inputenc}
\usepackage[T1]{fontenc}
\usepackage{mathptmx}
\usepackage{etoolbox}

\makeatletter
\def\@email#1#2{%
 \endgroup
 \patchcmd{\titleblock@produce}
  {\frontmatter@RRAPformat}
  {\frontmatter@RRAPformat{\produce@RRAP{*#1\href{mailto:#2}{#2}}}\frontmatter@RRAPformat}
  {}{}
}%
\makeatother
\begin{document}

\preprint{AIP/123-QED}

\title[PBML for mantle convection]{Physics-based machine learning for mantle convection simulations}
\author{Siddhant Agarwal}
 \affiliation{Institute of Space Research, German Aerospace Center (DLR), Berlin}
 \affiliation{Model-driven Machine Learning, Helmholtz-Zentrum Hereon, Geesthacht}
\email{siddhant.agarwal@dlr.de}

\author{Ali Can Bekar}
 \affiliation{Model-driven Machine Learning, Helmholtz-Zentrum Hereon, Geesthacht}

 \author{Christian H{\"u}ttig}
 \affiliation{Institute of Space Research, German Aerospace Center (DLR), Berlin}

 \author{David S. Greenberg}
 \affiliation{Model-driven Machine Learning, Helmholtz-Zentrum Hereon, Geesthacht}

 \author{Nicola Tosi}
 \affiliation{Institute of Space Research, German Aerospace Center (DLR), Berlin}

\date{\today}

\begin{abstract}
Mantle convection simulations are an essential tool for understanding how rocky planets evolve. However, the poorly known input parameters to these simulations, the non-linear dependence of transport properties on pressure and temperature, and the long integration times in excess of several billion years all pose a computational challenge for numerical solvers. We propose a physics-based machine learning approach that predicts creeping flow velocities as a function of temperature while conserving mass, thereby bypassing the numerical solution of the Stokes problem. A finite-volume solver then uses the predicted velocities to advect and diffuse the temperature field to the next time-step, enabling  autoregressive rollout at inference. For training, our model requires temperature-velocity snapshots from a handful of simulations ($94$). We consider mantle convection in a two-dimensional rectangular box with basal and internal heating, pressure- and temperature-dependent viscosity. Overall, our model is up to $89$ times faster than the numerical solver. We also show the importance of different components in our convolutional neural network architecture such as mass conservation, learned paddings on the boundaries, and loss scaling for the overall rollout performance. Finally, we test our approach on unseen scenarios to demonstrate some of its strengths and weaknesses.
\end{abstract}

\maketitle


\section{\label{sec:intro}Introduction}
\subsection{\label{sec:intro_otivation}Motivation}

Mantle convection plays a fundamental role in the long-term thermal evolution of rocky bodies, such as Earth, the Moon, Venus, Mercury, and Mars. Key processes that shape the evolution of a planet over billions of years, including its potential habitability, such as volcanism, tectonics, and magnetic field generation are all intricately related to how the silicate mantle and crust transport heat from the deep interior to the surface \citep[e.g.][]{schubert_book,breuer2015,tosi_book}. Mantle rocks behave as a highly viscous fluid over geological timescales in response to buoyancy forces induced by extreme temperatures and pressures. As such, mantle convection is modeled via a system of partial differential equations (PDEs) describing conservation of mass, momentum, and energy, which can be numerically solved using dedicated codes \citep[e.g.,][]{stagyy, zhong2008, huettig2013, aspect-doi-v3.0.0}. 

Initial conditions and several parameters for mantle convection simulations are poorly constrained and need to be extensively varied in parameter studies to match sparse observations from satellites, telescopes, and in-situ measurements \citep[e.g.,][]{Tosi2021}. Since simulations are computationally expensive, parameter studies are often carried out using simplified models based on scaling laws. These laws parameterize convective heat fluxes as a function of simulation parameters like the Rayleigh number and the rheological properties of rocks \citep[e.g.,][]{reese1998,dumoulin1999,deschamps2001,korenaga2010,thiriet2019,schulz2020}. 

Machine learning has started to overcome the limitations of scaling laws by incorporating more physics, such as temperature- and pressure-based dependence of the viscosity, and by predicting the mean temperature \citep{shahnas2020}, one-dimensional temperature profiles as a function of time \citep{agarwal2020}, as well as two-dimensional temperature fields \citep{agarwal2021b} in time. \citet{agarwal2021b} modeled the spatio-temporal evolution of the 2D temperature field in a Mars-like planet by first using convolutional autoencoders to compress the data, followed by training a recurrent algorithm to perform time-stepping in the latent space. On the one hand, this model is four orders of magnitude faster at inference than the numerical solver and is able to effectively capture the diverse flow patterns resulting from varying parameter combinations. On the other hand, the authors acknowledged some shortcomings as well: (1) the approach required $10,525$ simulations for training and evaluation, making it unfeasible to regenerate such large datasets on high-performance computing systems whenever the simulation setup is modified -- for example, when changing the domain geometry, the number of parameters, or the viscosity formulation; (2) the model did not predict other important variables such as pressure and velocity fields; and (3) it failed to accurately capture fine-scale features, such as small-scale downwellings arising from instabilities in the cold upper boundary layer. Recent studies, however, have begun to address this issue by developing models that can resolve such fine-scale features more accurately \citep{pathak2020usingmachinelearningaugment, wu2022learningacceleratepartialdifferential, yin2023continuouspdedynamicsforecasting}.

These limitations lead to the following question: \textit{Are machine learning models of mantle convection doomed to subpar predictions, despite training on thousands of simulations?} We tackle this question with a physics-based model that is trained on only $94$ simulations and is up to $89$ times faster than the numerical solver. 

\subsection{\label{sec:intro_pbml}Physics-based machine learning}

It is worth zooming out of the specific case of mantle convection to consider the wider efforts in physics-based machine learning for PDEs. \citet{pinn} is often cited as one of the seminal papers where automatic differentiation is used to evaluate the PDE terms, which are then incorporated into the loss function. Several follow-up works have proposed improvements, such as for overcoming spectral bias \citep{shishehbor_parametric_2024}, for better shock-capturing in transonic flows \citep{wassing2025}, for overcoming local minima in multi-loss objectives \citep{liu2025config}, and for multi-scale modeling \citep{wangmultiscalepinn}, to name a few. We refer to \citet{cuomo_scientific_2022} for a comprehensive review of these methodologies. 

While it is somewhat a matter of semantics, the term PINN (physics-informed neural networks) is generally associated with methods where the inputs to the neural network are coordinate-based points and automatic differentiation (AD) is used to calculate the partial derivatives of the outputs with respect to the inputs. However, convolutional neural networks (CNN) and graph neural networks (GNNs) have also been applied successfully in some cases where accounting for the spatial structure of the inputs is desirable. For example, \citet{wandel2020} use CNNs to predict flow variables and use a finite difference formulation in the form of convolutional kernels to calculate the partial derivatives instead of using AD. Notably, they do not use any data and solve the system of equations in the training phase. 

Hybrid methods have also gained popularity. \citet{Kochkov,alieva_2023,wang2024pcnet, um2021solverinthelooplearningdifferentiablephysics} augment coarse-grid solutions obtained from PDE solvers with a learned correction to account for high-resolution features. \citet{tompson2017} replace the expensive pressure projection step in a numerical solver with an unsupervised formulation to obtain divergence-free velocities for their incompressible Euler equations. The network is trained in the framework of several time-steps where their CNN block can be repeatedly called after each advection step, which helps establishing long-term stability. \citet{agarwal2025} could be considered a hybrid approach, in which learned one-dimensional temperature profiles serve as optimal initial conditions for a 2D mantle convection solver, allowing it to reach steady- or statistically-steady states $2.8$ times faster than with conventional initializations. Achieving this speedup at the cost of only $\sim$2 minutes of training time is appealing. However, this benefit is limited to scenario of the steady-state solution, rather than the full time-stepping process. The latter is crucial in thermal evolution simulations, where the system continue to evolve based on addition and removal of heat. Our new hybrid approach instead targets acceleration of the time-stepping itself, enabling faster simulation of the full temporal evolution.

\subsection{\label{sec:intro_contributions}Our approach}

We replace the most computationally expensive component of mantle convection simulations --  the solution of the mass and momentum conservation equations (i.e., Stokes problem) -- with a learned CNN that predicts velocities as a function of temperature and enforces mass conservation by design. The divergence-free velocities obtained from the CNN via a PyTorch model are then fed into the C++ numerical solver GAIA \citep{huettig2013} via a Python interface to perform a numerically-inexpensive advection-diffusion step. In this way, we are able to perform time-stepping without ever learning in time. As we will see later, the mantle convection PDEs provide a strong inductive bias on how to model in time and not accounting for it can make the learning task more challenging. Our contributions in this paper are as follows:

\begin{itemize}
    \item To the best of our knowledge, this is the first physics-based machine learning model in mantle convection.
    \item We introduce a scaling for learning velocity fields across several orders of magnitude.
    \item We use learned paddings on the boundaries for enhanced accuracy.
    \item We achieve stable predictions over tens of thousands of time-steps without learning in time.
    \item We evaluate this model on unseen scenarios such as thermal evolution at inference time and demonstrate the strengths and shortcomings of our approach.
\end{itemize}

We use a similar architecture to the one introduced by \citet{tompson2017}, but need more trainable parameters for our Stokes problem as opposed to their Poisson problem. In our case, pressure projection is not necessary, as the network conserves mass through constraints. This is similar to how \citet{wandel2020} enforce mass conservation. However, we adopt a data-driven approach for the momentum equation instead of a purely physics-based one. Not needing training data from the solver is attractive, but reaching the efficiency and accuracy of computational fluid dynamic codes with machine-learning-based solvers is an active area of research \citep{kups64227, wandel2025metamizer}. 
Our hybrid approach instead learns on high-quality samples from the solver. All the machine learning code developed and used is available here: \url{https://github.com/agsiddhant/PBML_Mantle_Convection}.

The paper is organized as follows. In Sec. \ref{sec-methods_eq}, we introduce the governing PDEs and the setup of the simulations. In Sec. \ref{sec-methods_dataset} we provide an overview of the dataset and discuss the scaling of velocities. We then present the architecture of the CNN used as a Stokes surrogate model in Sec. \ref{sec-methods_cnn} followed by two baseline models for comparison in Sec. \ref{sec-methods_baseline}. In Sec. \ref{sec-results} we discuss our main findings: training details (Sec. \ref{sec-res-train}); velocity predictions form the Stokes model (Sec. \ref{sec-res-vel}); time-evolution using a U-net (Sec. \ref{sec-res-unet}); performance of our Stokes model across different parameters during rollout (Sec. \ref{sec-results_parameters}); speedup analysis compared to the numerical solver and other baselines (Sec. \ref{sec-results_speedup}); ablations of key components, where we remove one component at a time to assess its impact on performance (Sec. \ref{sec-results_ablations}); and performance of our model on some unseen scenarios (Sec. \ref{sec-results_generalizability}). Finally, we conclude by summarizing the main findings of this paper.

\section{Methods}
\label{sec-methods}

\subsection{Mantle convection equations}
\label{sec-methods_eq}

The large-scale deformation of crystalline mantle rocks over geological time-scales is typically modeled as the dynamics of a viscous fluid with negligible inertia, leading to the so-called Stokes flow. We assume the Boussinesq approximation, whereby the flow is treated as incompressible and the only density variations considered are those due to temperature changes in the buoyancy force term \citep[e.g.,][]{schubert_book}. In non-dimensional form, the conservation equations of mass, momentum and thermal energy for a creeping fluid heated from below and from within in a 2D cartesian geometry read:
\begin{subequations}
\begin{align}
\nabla \cdot \boldsymbol{u} &= 0, \label{eq:mass}
\\
- \nabla p + \nabla \cdot \left( \eta \left( \nabla \boldsymbol{u} + (\nabla \boldsymbol{u})^{\textrm{T}} \right) \right) &= Ra \; T \; \boldsymbol{e}_y, \label{eq:momentum}
\\
\frac{\partial T}{\partial t} + \boldsymbol{u} \cdot \nabla T &= \nabla^2 T + Q .\label{eq:energy}
\end{align}
\end{subequations}
Here, $\boldsymbol{u}$ is the velocity vector with horizontal and vertical components $u$ and $v$, $p$ is the dynamic pressure, $T$ is the temperature, $\boldsymbol{e}_y$ is the unit vector in the vertical direction, $Q$ is the internal heating rate, $\eta$ is the dynamic viscosity, which depends on temperature and depth (i.e., hydrostatic pressure) as
\begin{equation}
\eta(T,y) = \exp\left( -\log(\gamma) T  + \log(\beta) (1-y) \right),
\label{eq:eta}
\end{equation}
where  $y$ is the height ($y = 0$ at the bottom of the domain and $y = 1$ at the top), and the parameters $\gamma$ and $\beta$ denote the maximum viscosity contrasts due to temperature and depth, respectively. $Ra$ is the Rayleigh number:
\begin{equation}
    Ra=\frac{\rho^2 \; c_p \; g \; \alpha \; \delta T \; D^3}{\eta_0 \; k},
    \label{eq:Ra}
\end{equation}
where, where $\rho$ is the density, $c_p$ is the heat capacity, $g$ is the gravitational acceleration, $\alpha$ is the coefficient of thermal expansion, $\delta T$ is a temperature scale, $D$ is the height of the domain, $k$ is the thermal conductivity, and $\eta_0$ is a reference  viscosity. $Ra$ is thus the ratio of buoyancy forces due to thermal expansion that drive convection to thermal diffusion and viscosity that inhibit it.

As can be seen from Eq. \eqref{eq:momentum}, for a fluid with negligible inertia, the velocities have no memory of previous time-steps and are fully determined by the temperature field and the resulting viscosity field. These velocities then determine how the temperature field is advanced in time in Eq. \eqref{eq:energy} via advection and diffusion, where heat sources and sinks can be optionally added and can even vary in time, such as in the case of radioactive decay, relevant for planetary interior evolutions.

We solve the above equations in a 2D rectangular box with an aspect-ratio of four. All boundaries are impermeable (zero normal velocity), and free-slip (zero shear stress). The top and bottom boundaries are isothermal, while the sidewalls are insulating. This translates into the following set of Dirichlet and Neumann boundary conditions:

\begin{itemize}
  \item \textit{Left:} $u=0$, $\displaystyle\frac{\partial v}{\partial x} = 0$, $\displaystyle\frac{\partial T}{\partial x} = 0$
  
\item \textit{Right:} $u=0$, $\displaystyle\frac{\partial v}{\partial x} = 0$, $\displaystyle\frac{\partial T}{\partial x} = 0$
  
 \item \textit{Top:} $v=0$, $\displaystyle\frac{\partial u}{\partial y} = 0$, $T = 0$
  
\item \textit{Bottom:} $v=0$, $\displaystyle\frac{\partial u}{\partial y} = 0$, $T = 1$

\end{itemize}

\subsection{Dataset and scaling}
\label{sec-methods_dataset}

We use the same dataset as in \citet{agarwal2025}, where the simulations attain a statistical steady-state after typically advancing the solution for tens of thousands of time-steps. While discovering steady-states of the mantle convection simulations is of great interest to the mantle convection community, we note that our model enables fast time-stepping and can be applied to evolution scenarios as well. In fact, we will apply the model learned on this steady-state dataset to an evolution scenario later, where the internal heat source is no longer constant, but decays in time.

The dataset is generated with the finite volume code GAIA on a uniform grid of $128 \times 506$ cells. Mass and momentum conservation are solved with the MUMPS direct solver \citep{MUMPS:1, MUMPS:2}, while the energy equation is solved with an iterative solver. The dataset consists of $128$ simulations, of which $94$ are used for training, $16$ for cross-validation, and $18$ for testing - all chosen randomly. Three simulation parameters are randomly sampled from a uniform distribution to generate the dataset: $Q$ (between 0 and 10), $\beta$ (between 1 and 100), and $\gamma$ (between $10^6$ and $10^{10}$). The Rayleigh number in Eq. \eqref{eq:momentum} is set to 1. The vigor of convection is influenced through an effective Rayleigh number through the temperature- and pressure-contrasts. The range of parameters yields effective Rayleigh number values ranging from $10^4$ (when $\gamma=10^6$ and $\beta=100$) to $10^{10}$ (when $\gamma=10^{10}$ and $\beta=1$).

\begin{figure*}
\centerline{\includegraphics[width=\textwidth]{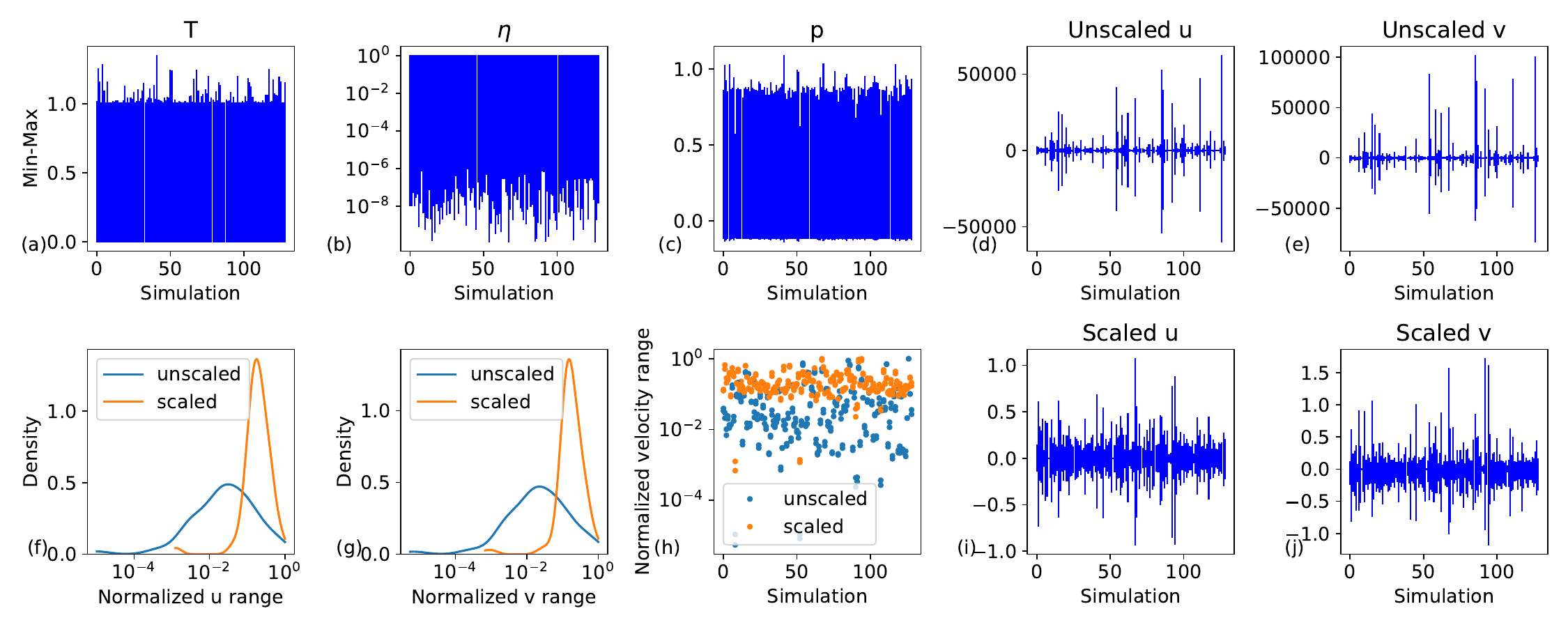}}
\caption{The ranges of different fields in the simulations (a-e). Scaling the velocities brings the ranges across different simulation parameters closer to each other (f,g). For ease of visualization, we normalize the $u$ and $v$ range for the scaled and unscaled cases by dividing them by the maximum (f,g,h). }
\label{fig-scalings}
\end{figure*}

In Fig. \ref{fig-scalings}, we visualize the range of each field by drawing a line from the minimum to the maximum of each simulation. Temperature (a), viscosity (b) and pressure (c) are well-behaved. For viscosity, we can take the $log_{10}$ and divide by $8$ so that it varies from approximately $-1$ to $0$. However, the minimum and maximum velocities (d, e) span different orders of magnitudes -- for some simulations these span large ranges, whereas for the others these remain much more limited. There are two issues with the standard normalizations such as min-max scaling, mean-standard deviation scaling, and log scaling. First, we do not want to shift the mean of the velocities, but only divide by some constant both sides of Eq. \eqref{eq:mass}, so that mass conservation is not violated. Second, dividing by the overall maximum of the velocities would still leave simulations with values spanning a significantly smaller range compared to the wide ranges spanned by other simulations. As the magnitude of the velocity field is correlated with the simulation parameters, we fit a linear regression model through the maximum values of $u$ and $v$ as function of the simulation parameters and obtain the following scaling (shown here up to two significant digits):
\begin{equation}
    \text{velocity scaler} = 5\, e^{0.18 Q}\,\, \gamma^{0.43}\,\, \beta^{-0.46} 
    \label{eq:velocityscaling}
\end{equation}
Dividing the velocity fields by the above factor brings the simulations within a more homogeneous range. This is shown in Fig. \ref{fig-scalings}(f) for $u$ and Fig. 1(g) for $v$, where we compute the range as maximum minus minimum value of velocities for all the time-steps in a simulation. We normalize the ranges by the maximum values to facilitate comparison of scaled and unscaled velocities. When unscaled, there is a very small difference between the maximum and minimum value of velocities from each simulation. However, by scaling, we push most of the simulations to have a higher range (Fig. \ref{fig-scalings}h) by one order of magnitude or more. 

Our CNN thus learns on scaled velocity fields, which we can simply multiply by Eq. \eqref{eq:velocityscaling} to recover the actual, unscaled magnitudes at inference time. This technique was motivated by our experience with a simpler set of simulations, where only $Ra$ was varied and a uniform viscosity was used. In this case, velocities and pressure can still vary several orders of magnitude, but the magnitudes vary approximately linearly with $Ra$. Scaling the fields by $Ra$ enabled us to obtain good predictions. Thus, this approach is a multi-parameter extension with some modifications: viscosity is no longer constant and $Ra$ is always $1$, but different parameters drive the scales now.

\subsection{Machine learning model}
\label{sec-methods_cnn}

\subsubsection{Convolutional neural network}

\begin{figure*}
\centerline{\includegraphics[width=\textwidth]{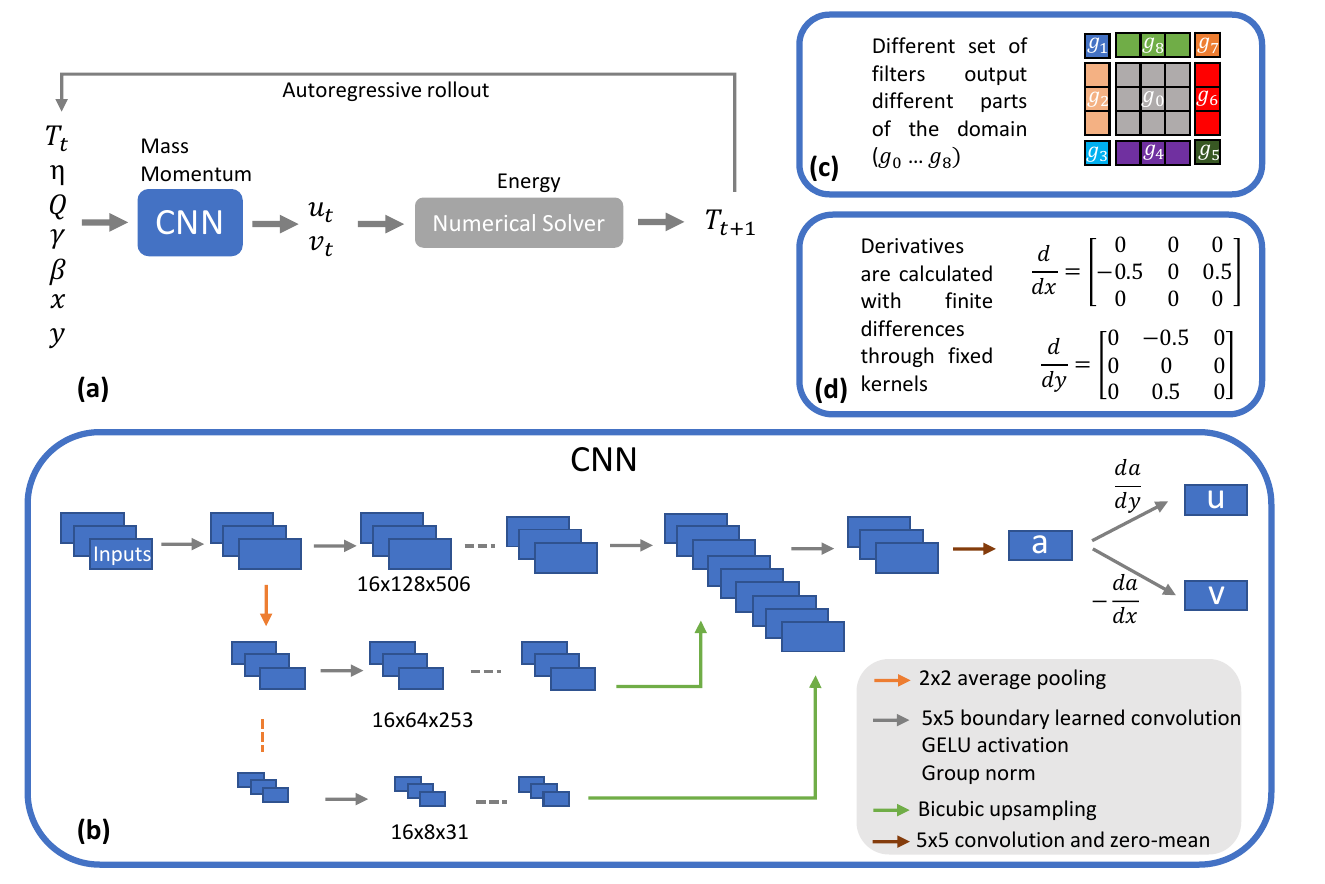}}
\caption{Our hybrid physics-based model for time-stepping. (a) We use a data-driven approach to model the primary computational bottleneck in mantle convection simulations: solving the Stokes equation. The velocities predicted by the convolutional neural network (CNN) can be fed to the numerical solver for time-stepping. (b) The CNN architecture is inspired by \citet{tompson2017} for modeling solutions of linear systems as a function of the right-hand-side of the equation. (c) Instead of standard padding methods (such as zero paddings), we use boundary-learned convolutions. A unique set of filters is responsible for predicting each corner and each edge as well as the interior domain. (d) Mass conservation is enforced by predicting a field whose curl yields divergence-free velocity components $u$ and $v$.}
\label{fig-method}
\end{figure*}

CNNs have been widely adopted in a variety of machine learning models for PDEs on uniform grids. As shown in Fig. \ref{fig-method}a, the CNN acts as a surrogate model for the velocities induced by the temperature field. The main input to the CNN is the temperature field, but we further enrich it with the viscosity field calculated via Eq. \eqref{eq:eta}. We also add coordinates of the numerical grid as is sometimes done in coordinate-based neural networks \citep{serrano2023operator, catalani_neural_2024} and graph neural networks \citep{graphcast,wei2024}. Finally, even though the simulation parameters $\beta$ and $\gamma$ are already contained in the viscosity field and $Q$ is irrelevant for Eqs. \eqref{eq:mass}--\eqref{eq:momentum}, we explicitly pass these as inputs to the CNN, as these are used to scale the velocities (Eq. \eqref{eq:velocityscaling}). The predicted velocities are then fed to the numerical solver to advect the temperature field. In this way the model can be used autoregressively at inference time. We use the same architecture as \citet{tompson2017} (Fig. \ref{fig-method}b), which was designed for modeling linear algebaraic systems. The input is first embedded into an abstract representation with more channels than the number of input fields. We downsample this representation five times by a factor of $2$ using average pooling. This allows the network to process features of different scales at different levels -- the receptive field of the coarsest level spans almost the entire domain. This is important for PDEs where one point in input can impact any other point in the output. At each resolution, the representation is independently processed with a series of convolutions before being upsampled via bicubic interpolation to the original resolution and concatenated. The upsampled features are concatenated instead of summed, allowing the network to learn how to combine this information through another series of convolutions. This differs slightly from the original U-net architecture where the information is downsampled and upsampled through an autoencoder-style architecture with skip connections between the encoder and decoder part at each resolution \citep{ronneberger2015u}. U-nets favor learning incremental updates to the inputs due to skip connections at each level. This is more sensible when learning time-stepping between states, whereas the architecture of \citet{tompson2017} seems more efficient for learning output variables (e.g., $u, v$) as a function of different input variables (e.g., $\eta, T$). 

\subsubsection{Mass conservation}

As in \citet{wandel2020}, we enforce mass conservation in our incompressible flow by predicting a vector potential $\textbf{a}$, whose curl provides divergence-free components of velocity. Working in two dimensions, we need to predict only one component of $\textbf{a}$, $a_z$, or simply, $a$:
\begin{equation}
    u =  \frac{\partial a}{\partial y}, \;\;\;\;\; v = -\frac{\partial a}{\partial x}
    \label{eq:vectorpotential}
\end{equation}
The derivatives are calculated as central finite differences using fixed convolutional filters (see Fig. \ref{fig-method}d). These are conveniently carried out on the GPU, as is all the training of the CNN. As in \citet{wandel2020}, we subtract the mean of the output from $a$ to keep the values bounded. We multiply $a$ with an empirically determined value of $10$ to help the network output values not stray too far beyond $-1$ and $1$. Different values of $dx$ only serve as a multiplicative factor for Eq. \eqref{eq:mass} that have little bearing on mass conservation in this case with a uniform grid. This is also the reason behind the form of the velocity scaling of Eq. \ref{eq:velocityscaling}: divergence-free network velocities multiplied by this factor remain divergence-free. In practice, with single precision training of the network, multiplying the velocities with large scaling factors might violate mass conservation beyond an acceptable threshold. Hence, we train with double precision, leaving further optimization of training precision for future work. 

Finally, we note that as one calculates the derivatives of $a$ with respect to $x$ and $y$, the resulting velocity fields are smaller in their respective dimensions by two layers of points. We proceed with padding these predictions as per the necessary boundary conditions: we pad with the exact value for Dirichlet boundary conditions and copy the value of the adjacent cell where Neumann boundary conditions are zero. This inevitably makes the mass conservation inexact at the boundaries. Therefore, we compute Eq. \eqref{eq:mass} just at the boundaries, and plug it into the loss function during training as a soft constraint, while, on the interior domain, mass conservation is already satisfied (down to double machine precision). An alternative to this would be to pad the network output $a$ with either learned values or values derived so that the boundary conditions would be satisfied. However, we were unable to learn a compatible padding. Setting up and solving a linear system would also be an option. Either way, it is not clear if a formulation of $a$ exists that can satisfy these exact boundary conditions. This highlights one of the challenges in physics-based machine learning, namely learning in a consistent manner with the data from the PDE solver. It is also worth noting that, no matter what values we prescribe at the boundaries, the solver will overwrite them in exactly the same way as we do by either assigning an exact value for a Dirichlet condition, or by copying the adjacent value for a Neumann condition. We later assess the effectiveness of a hard constraint vs. a soft constraint in rollout performance -- in the latter, the CNN predicts the velocities directly and mass conservation equation calculated on these velocities is plugged into the loss function. 

\subsubsection{Learned boundary paddings}

When convolving at the edges of the domain, the input to each layer must be padded if one wishes to maintain the original spatial dimensions after convolution. After a lot of experimentation, we found that typically used paddings such as zeros and replicate (i.e., copying the adjacent value) are detrimental for accuracy on the edges. This shortcoming became evident upon advecting the temperature field with the predicted velocities:  errors on the boundaries would give rise to artificial ``bubbles'' of hot material that would rapidly rise and destabilize the solution. \citet{cnnpadding} showed some advantage of using a spatial context of $0$ or $1$ to help CNN learn on the boundaries, but we found it to be insufficient when combined with replicate padding. 

We employed the ``boundary learned convolution'' from \citet{innamorati_learning_2020}, which learns the padding on the boundaries itself by decomposing the domain and learning a separate set of filters responsible for predicting the interior domain as well as all the edges and corners (see Fig. \ref{fig-method}c). The learned filters are applied to the spatial context of each subdomain to reproduce the original width and height. The intuition here is that the fields such as temperature and velocities, but also images in general, might have non-trivial yet smooth spatial extrapolations. In this case, adding zeros might create artifacts that the CNN has to learn to overcome. Using a unique set of filters for each subdomain increases the count of trainable parameters by a factor of $9$. With the exception of the interior filter, which is applied to the entire domain and produces $g_0$, all the ``extra'' filters are applied to significantly smaller subdomains (edges and corners) and therefore do not significantly increase the overall time complexity. Consequently, a network with same total parameter count but regular convolutions would be slower. 

\subsubsection{Loss function}

We use mean absolute error (MAE) between the true and predicted fields as our loss function. Despite the scaling described in Eq. \eqref{eq:velocityscaling}, the velocities can still vary significantly. In particular, the high velocity values will now be smaller due to the scaling. Therefore, to ensure that, within each batch, proper weight is given to the lower velocity magnitudes, we scale the loss by dividing it by the range of each example in the batch. We clip this loss scaling at $10$ to avoid creating too much imbalance between the examples:
\begin{align}
S_{\text{norm}} &= \text{clip} \left(\frac{1}{\max(x_{\text{true}}) - \min(x_{\text{true}})}, 1, 10 \right). \label{eq:loss_scaling}
\end{align}
We further add a factor of $10$ on the boundaries to penalize errors there more heavily than in the interior of the domain: 
\begin{align}
S_{\text{bc}} &= 
\begin{cases} 
1+10, & \text{on boundaries} \\
1, & \text{otherwise}.
\end{cases} 
\end{align}
These values are somewhat arbitrary and one could certainly optimize them further at the cost of more computational resources. However, we perform an ablation study to assess the impact of removing the loss scaling terms. The loss is calculated as: 
\begin{align}
L(x_{\text{true}}, x_{\text{pred}}) &= 
\begin{cases} 
\displaystyle\frac{1}{N} \sum \left| (x_{\text{true}} - x_{\text{pred}}) \cdot S_{\text{norm}} \cdot S_{\text{bc}} \right|,  & \text{if scale loss}\\
\displaystyle\frac{1}{N} \sum \left| (x_{\text{true}} - x_{\text{pred}}) \right|,  & \text{otherwise}. \\
\end{cases} 
\end{align}

We add mass conservation to the loss function if mass is conserved as a soft constraint. If the curl-based formulation is used, we only use the mass conservation calculations on the boundaries. The mass conservation term is calculated as:
\begin{align}
L_{\text{mass}} &= \frac{1}{N} \sum \left | \frac{\partial u}{\partial x} + \frac{\partial v}{\partial y}\right |
\end{align}
Finally, we found that the central-differences-based curl formulation produced some oscillations in the $y$ direction for $u$ and in the $x$ direction for $v$. Applying Gaussian filtering to $a$ did not seem to reduce these artifacts. Instead, we learn not only on the velocities, but also on their derivatives. The intuition behind this is that artifacts become even more prominent when their derivatives are calculated so they can be effectively penalized:
\begin{align}
L_{\text{deriv}} &=  \frac{1}{N} \sum \left | \frac{\partial u_{\text{true}}}{\partial y} - \frac{\partial u}{\partial y}\right | +  \frac{1}{N} \sum \left | \frac{\partial v_{\text{true}}}{\partial x} - \frac{\partial v}{\partial x}\right | \label{eq:loss_derivative}\\
\end{align}
The overall loss term can now be calculated as:
\begin{align}
L_{\text{total}} &= 
\begin{cases} 
\displaystyle\frac{L_u + L_v + L_T}{3}, & \text{if } T \text{ is predicted} \\
\displaystyle\frac{L_u + L_v}{2}, & \text{otherwise}
\end{cases} \\
L_{\text{total}} &\mathrel{+}= 
\begin{cases} 
\displaystyle\frac{1}{N} \sum L_{\text{mass}}, & \text{if soft constraint} \\
\displaystyle\frac{1}{N_\text{bc}} \sum L_{\text{mass, bc}}, & \text{if curl-based constraint}
\end{cases} \label{eq:mass_loss}\\
L_{\text{total}} &\mathrel{+}= L_{\text{deriv}}, \quad \text{if derivative loss term is included}
\end{align}

\subsection{Baselines}
\label{sec-methods_baseline}

We consider two different prediction models, each with a specific purpose.

\subsubsection{Numerical solver with sub-optimal settings}
\label{sec-methods_baseline_solver}

The velocities predicted by the CNN are relatively accurate but still exhibit some discrepancies compared to the ground truth solution obtained from our numerical PDE solver. Given that advection is still possible using these slightly erroneous velocities, it is natural to ask how an imperfect solver would perform in terms of both accuracy and computational speedup. To this end, we consider as first option solving the mass and momentum equations only every $100$ time-step. 
If our model outperforms this benchmark, it would suggest that we can produce meaningful temporal interpolations, despite the training data being available only at every 100th time step. Notably, 100 was also the number of skips required to achieve a speedup comparable to that provided by the CNN. On one hand, this baseline is somewhat unfair to our best model, since we could likely skip several steps before reaching the error level of the solver with 100 momentum skips. On the other hand, the suboptimal solver is readily available without the need for data or training.

As a second option, we consider an iterative solver running on the same GPU used for CNN inference. As the iterative solver is slower than the direct solver, we use an under-relaxation factor of $0.99$ to still achieve some speedup with respect to the direct CPU baseline and to evaluate its performance against the CNN.

\subsubsection{U-net for learning in time}
\label{sec-methods_baseline_unet}

For our second model, we choose a simple U-net architecture to explore how well time-stepping can be learned. In this case, we no longer rely on the numerical advection-diffusion solver, but instead learn to predict from the velocities and temperature at a given time-step based on the state at the previous time-step, the grid coordinates, the viscosity, and the time-step $dt$.

To keep the architecture as close as possible to our Stokes surrogate model, we downsample the original resolution $5$ times, and upsample it back to the original resolution using the same operations. At every level, the encoder and decoder each contain $3$ convolutional layers. Information is passed at each level from encoder to decoder through skip connections via concatenation of channels. We train four different versions: 

\begin{enumerate}
    \item U-net-1: We match the overall parameter count of the network to our Stokes surrogate model without our proposed techniques, i.e. boundary-learned convolution and curl-based mass conservation.
    \item U-net-2: We match the inference time of a single time-step from the network to that of a single $u$, $v$ prediction of the Stokes model without our techniques. 
    \item U-net-3: We match parameter count and use our techniques.
    \item U-net-4: We match inference time and use our techniques.
\end{enumerate}

\section{Results and discussion}
\label{sec-results}

\subsection{Training details for the Stokes model}
\label{sec-res-train}

After some trial-and-error of different hyperparameters in our Stokes surrogate model, we found that $16$ $5\times5$ filters, processing features $6$ times at each level, provided a good balance between speed and accuracy for predicting the flow velocities. Unless otherwise stated, all results are presented with this architecture. We train on a single GPU as we use a small batch size of $16$, which has been shown to improve generalization and contribute to stable rollouts, as observed in \citet{agarwal2021b}. We aim for $150$ epochs of training and reduce the learning rate by a factor of $0.5$ every $20$ epochs. In practice, not all networks reach $150$ epochs as we sometimes terminate training early to manage training costs. The network we present the results for and refer to as ``our best model'' required about $5$ days of training time to reach $80$ epochs on an NVidia V-100 GPU. We note that we do not train on all the snapshots available (training set: $217,813$, cross-validation (cv) set: $44,049$, test set: $42,872$). For each simulation, we pick the first $200$ snapshots and then randomly select up to $800$ more as large portions of the time-series are in a statistical steady-state and do not have as much variability as the earlier time-steps. In this way, we are able to trim the data by $2/3$, which allows us to load all the data in the CPU memory and train different networks in parallel. This means that we use only $1/300$ of the data with respect to the original temporal resolution of the solver. One could further speedup training by processing larger batches on multiple GPUs and evaluate generalization performance. Finally, training in single or mixed precision could also help, if mass conservation is still satisfactorily satisfied if velocities with different orders of magnitude can be learned effectively.

\subsection{Velocity predictions with the Stokes model}
\label{sec-res-vel}

\begin{figure*}
\centerline{\includegraphics[width=\textwidth]{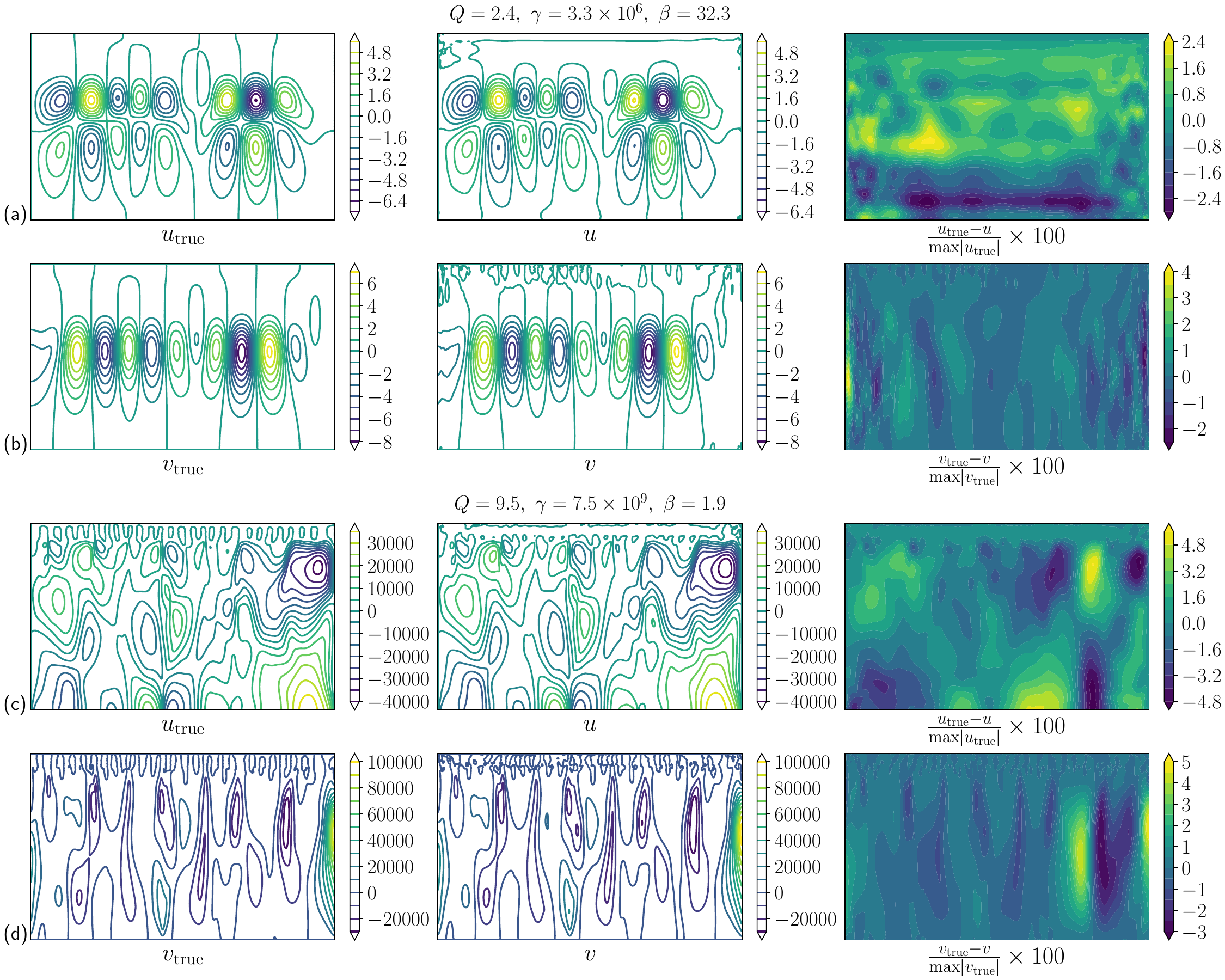}}
\caption{Examples of velocity predictions compared to ground truth from the direct solver for two different unseen simulations (a,b) and (c,d) characterized by widely different parameters leading to relatively low (a,b) and high velocities (c,d). A single model is able to learn across four orders of magnitude. The prediction error is divided by the absolute maximum value and displayed as a percentage error in the right column.}
\label{fig-velocities}
\end{figure*}

In Fig. \ref{fig-velocities}, we show examples of $u,v$ predictions from two different test-set simulations. The errors reach up to $5\%$ of the maximum magnitude of the true velocity in the second sample. This is quite high for a scientific machine learning study. However, we must keep in mind that a single model is learning on velocities across a few orders of magnitude from a handful of simulations. We deliberately picked two extreme cases from our test set here to demonstrate learning across different magnitudes. This error might also not be the most informative metric, as our goal is to advect the temperature using these velocities. While fitting a larger network could help reducing the errors, it might come at the cost of a lower generalization capability to unseen simulations, potentially leading to unstable rollouts. A larger network would also be slower at inference time. On average, we find that our best model predicts velocities $u_t,v_t$ that are approximately $14$ times more accurate than if one were to simply take the velocities at the last available time-step ($u_{t-100}$, $v_{t-100}$) from the solver. 

\subsection{U-net predictions for time-stepping}
\label{sec-res-unet}

\begin{figure*}
\centerline{\includegraphics[width=1\textwidth]{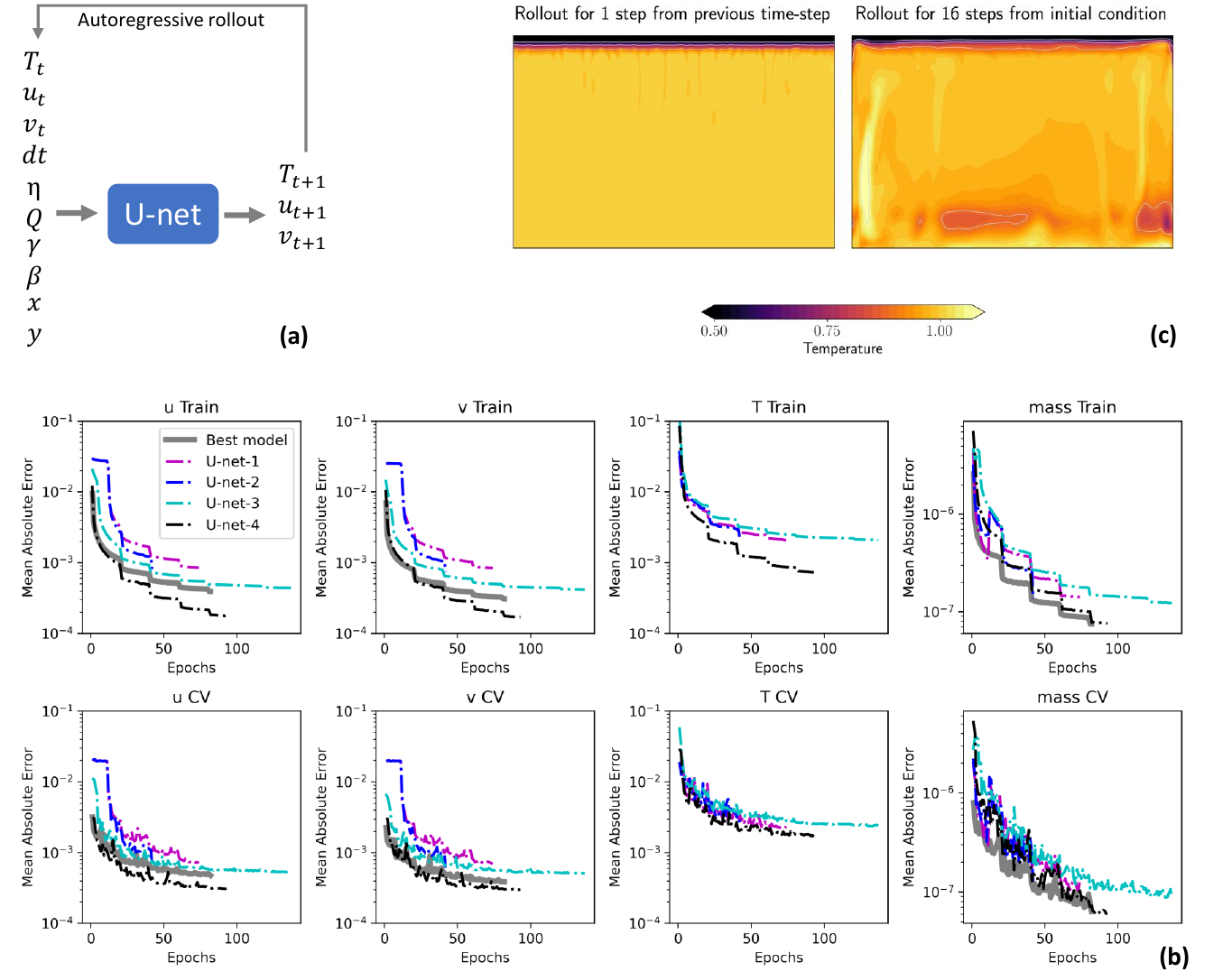}}
\caption{(a) U-net schematic for predicting from state at time $t$ to state at $t+1$. (b) Four different U-nets are trained to evaluate if different parameter counts and components like boundary-learned convolution improve the prediction accuracy. These are plotted alongside our best Stokes surrogate model that predicts velocities (solid grey lines). (c) When rolling out autoregressively with the best performing network (dashed black lines), the predictions diverge within $16$ steps.}
\label{fig-unet}
\end{figure*}

As described in Sec. \ref{sec-methods_baseline_unet}, we train four different U-nets as baselines to compare against our method. We plot the training curves in Fig. \ref{fig-unet} alongside our best model. Our techniques of boundary-learned convolution and loss scaling improve the loss attained during training compared to their counterparts without these techniques. Without them, we observe some artifacts where the network seems to mix up the fields in certain regions near the top boundary. This could be because the input to the network differs from the output. This highlights the challenge of learning with U-nets beyond incremental updates to the input as the system states evolve in time. However, we are able to overcome these artifacts with the boundary learned convolution and possibly also due to loss scaling. In fact, U-Net-4 (the same inference time network with our techniques) outperforms our best Stokes surrogate model in terms of training and validation loss on velocities (Fig. \ref{fig-unet}b). 

Nevertheless, as we rollout autoregressively during inference time with our best U-net on a mere 16 steps ($\sim 16\times100$ simulation time-steps), the solution quickly diverges (Fig. \ref{fig-unet}c). Several tricks have been suggested for overcoming this accumulation of errors during rollout \citep[e.g.,][]{lippe2023pderefiner} such as training on multiple time-steps. As training on multiple time-steps would increase the memory requirements by tracking a longer graph, the so-called ``pushforward'' trick does not track gradients in the rollout during training except for the last time-step. In this way, the network learns to overcome non-trivial errors that tend to accumulate. This, however, still comes at the cost of increased training time and intuitively makes the learning task more challenging as the network has to learn to de-noise the input in addition to learning a relation between the left hand side and the right hand side of a system of PDEs.

Thus, a large array of machine learning methods remains to be explored for directly time-stepping in mantle convection besides the recurrent learning approach of \citet{agarwal2021b}. These methods are attractive for their significant speedups, ability to process several simulations at once in batch, and end-to-end differentiability with simple models built in open-source frameworks like PyTorch. Nevertheless, we would like to offer an alternate perspective here. From Eq. \eqref{eq:momentum}, we know that the velocities at a given time are dependent on temperature, but are not directly a function of velocities at the previous time-step -- this would not be the case for non-linear, strain-rate dependent rheologies \citep[e.g.,][]{schulz2020}. Learning new velocities as a function of the previous velocities can thus introduce questionable correlations in our model. Although the best U-net outperforms the Stokes surrogate model in single-step velocity predictions, it fails to roll out, perhaps because it is conditioned on and sensitive to previous velocities. A more physics-based improvement to this U-net would be to learn two different networks: one that learns  $u_t,v_t$ as a function of $T_t$ (like our Stokes surrogate) and then another network that learns $T_{t+1}$ as a function of $T_t$, $u_t$, $v_t$. These could be optimized together or separately, but, either way, it would be interesting to explore how well the advection-diffusion operation can be learned. Given that we only have access to time steps sampled at intervals of 100, it remains an open question what the most effective approach is for learning this operator. 

Would the network (say $f$) be more accurate if we trained with a rollout of $100$ as $f(f({f(t_{0})_0})_1...)_{99}$, or would we be able to learn the mapping $f(t_0)_{99}$? This small thought experiment could be further extended to directly learning $T_{t_{99}} = f(T_{t_{0}})$, but would any features in the network look like velocity fields without explicitly encoding Eq. \eqref{eq:energy} and what would be the data-efficiency of such a network? The answer to the first question is no, and to the second, probably poor. \citet{agarwal2021b} showed that it is possible to skip velocities and learn this direct mapping from one temperature field to the next $200^{th}$, but at the cost of $10,000$ training simulations. One can seemingly tradeoff data-efficiency and causality for faster time-stepping. For fields where various simulation setups exist in the form of geometries, heat sources, viscosity models, parameter variations, phases transitions and melting models, data-efficiency can be an important consideration. This can be further viewed through the lens of \citet{buitrago2025on} who demonstrate the benefits of learning on multiple time-steps in the past through a ``memory'' for systems with partially observable (e.g., missing spatial points) or corrupt states (e.g., noise or uncertainty). \citet{agarwal2021b} benefited from using $16$ time-steps as inputs to their Long-Short-Term-Memory network, possibly because they learn only on $T$, and the missing velocities as well as the skipped time steps can be seen as an extreme case of missing observables.

\subsection{Rollout performance for different parameters}
\label{sec-results_parameters}

\begin{figure*}
\centerline{\includegraphics[width=\textwidth]{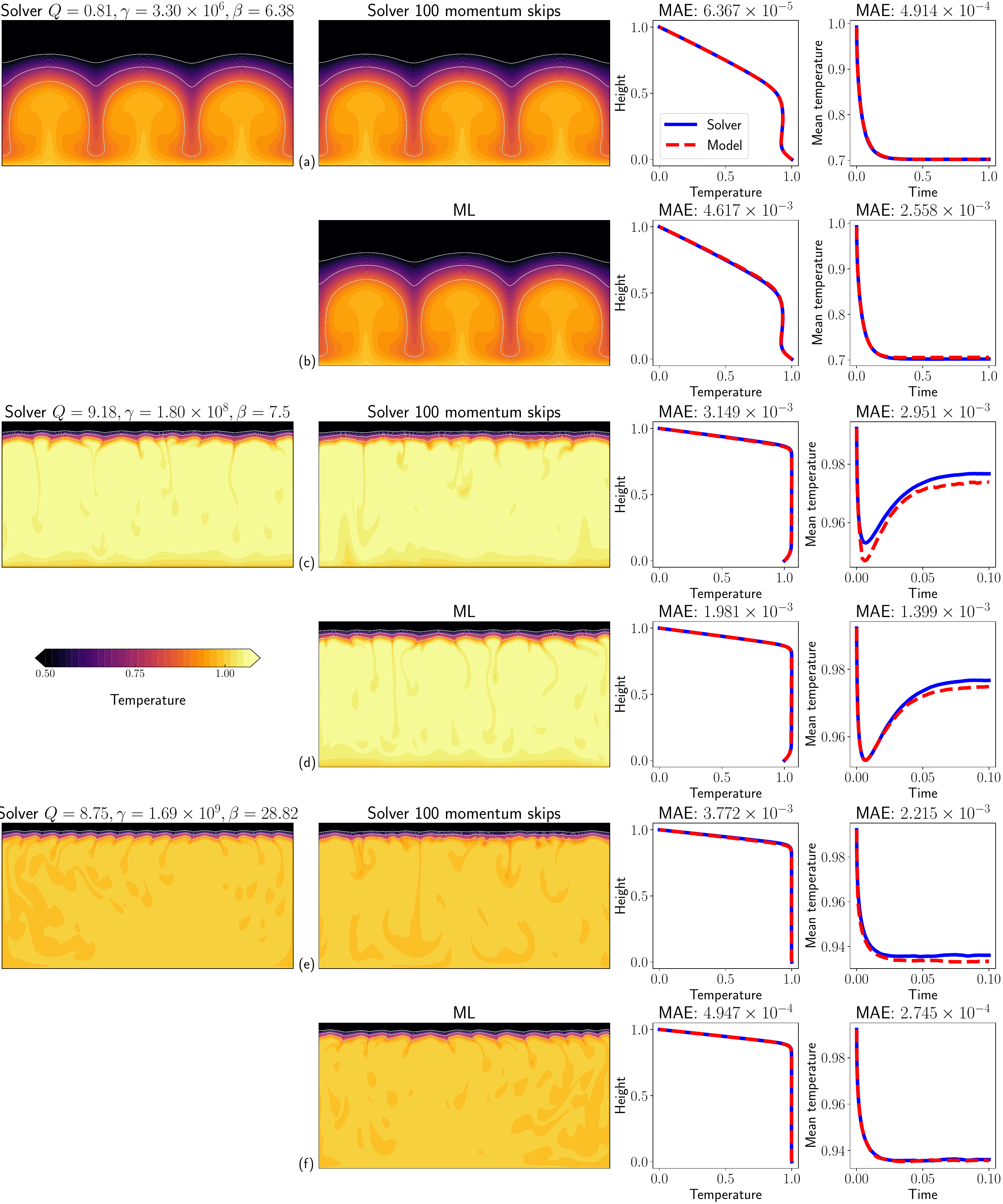}}
\caption{Evaluation of rollout performance on three qualitatively different simulations (a,b), (c,d) and (e,f). For each simulation, we plot the ground truth in the first column, followed by the prediction (either baseline of $100$ momentum skips, or the ML model) in the second, the horizontally-average temperature profile in the third, and the mean temperature as a function of time in the fourth.}
\label{fig-parameters}
\end{figure*}

In contrast to direct time-stepping with the U-net, we are able to produce a stable rollout for tens of thousands of time-steps using our hybrid approach. The velocities predicted by our Stokes model are used to advance the temperature field using the numerical advection-diffusion solver in GAIA. To make all runs comparable, we run the advection-diffusion solver with a Courant number of $1$, although the implicit solver allows for higher numbers. 

Fig. \ref{fig-parameters} shows three qualitatively different unseen simulations from the test set. These include sluggish convection with low frequency characteristics (Fig. \ref{fig-parameters}a,b), vigorous convection with higher frequency structures and a thinner upper thermal boundary layer (Fig. \ref{fig-parameters}e,f), and an intermediate case in Fig. \ref{fig-parameters}c,d, but with a non-monotonic time-series where the mean temperature  initially decreases and then increases before stabilizing. With the exception of the sluggish convection case, our model outperforms the suboptimal numerical solver in matching convection patterns such as pronounced undulations in the shape of the stagnant lid as well as in the horizontally-averaged temperature profile. The momentum skips and the resulting errors in the velocities are especially detrimental during the initial transient phase of vigorous convection, causing the intermediate case to miss the trough in the mean temperature curve of row (c) in Fig. \ref{fig-parameters} - this is undesirable in thermal evolution scenarios, where the history of the planet can contain clues to inferring certain parameters and thus requires accurate time-series modeling. 

For the sluggish case, the predictions by our model still capture the correct pattern of convection (\ref{fig-parameters}b), but the mean quantities fare slightly worse than the suboptimal solver. This is likely a consequence of the loss scaling of Eq. \eqref{eq:loss_scaling} as it assigns up to $10$ times more weight to more vigorously convecting simulations. Reducing the maximum clipping value could help mitigate this issue and result in a more favorable balance, but this issue ultimately highlights the challenge of learning a single model on disparate scales. 

\begin{figure*}
\centerline{\includegraphics[width=\textwidth]{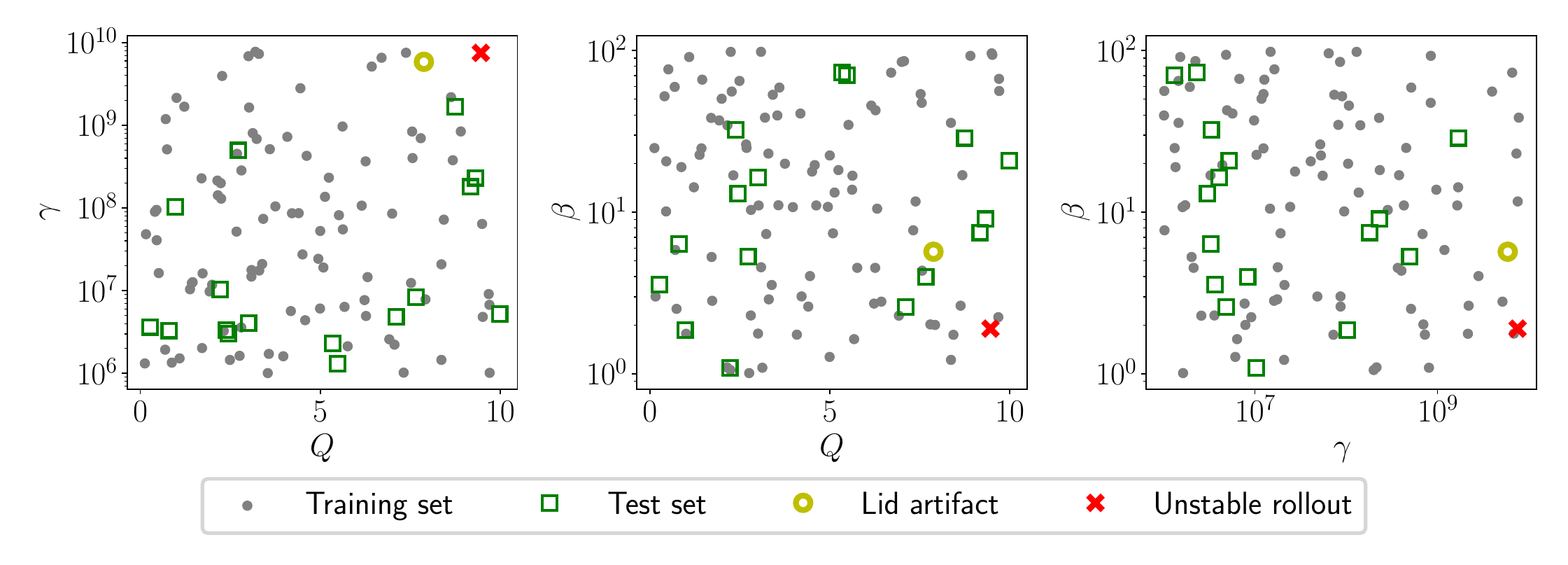}}
\caption{Parameter space of the training and test sets. We achieve stable rollout on all the simulations in the test set, except when extrapolating in the high $Q$-high $\beta$ space (e.g., yellow circle or red cross).}
\label{fig-stability}
\end{figure*}

We also found that the prediction accuracy diminishes significantly when extrapolating in the parameter space, highlighting another drawback of our approach. Again, the suboptimal solver performs better than our model when extrapolating. We were able to successfully run $16$ out of $18$ simulations in our test set. One run produced non-physical lid structures while another diverged after a few thousand steps. We mark these cases in Fig. \ref{fig-stability} and notice that this region of the parameter space is quite challenging. This is not only true from a physical standpoint of high $Q$ and $\gamma$ driving vigorous convection, but also from a data point of view, as the failed case (red cross) is clearly beyond our training data range and the nonphysical case (yellow circle) sits right on the edge of the $Q$-$\gamma$ envelope. In the future, we would like to add some additional training points at the exact corners of the parameter space, i.e. $8$ more training simulations.

\subsection{Comparison of speedups}
\label{sec-results_speedup}

\begin{figure*}
\centerline{\includegraphics[width=\textwidth]{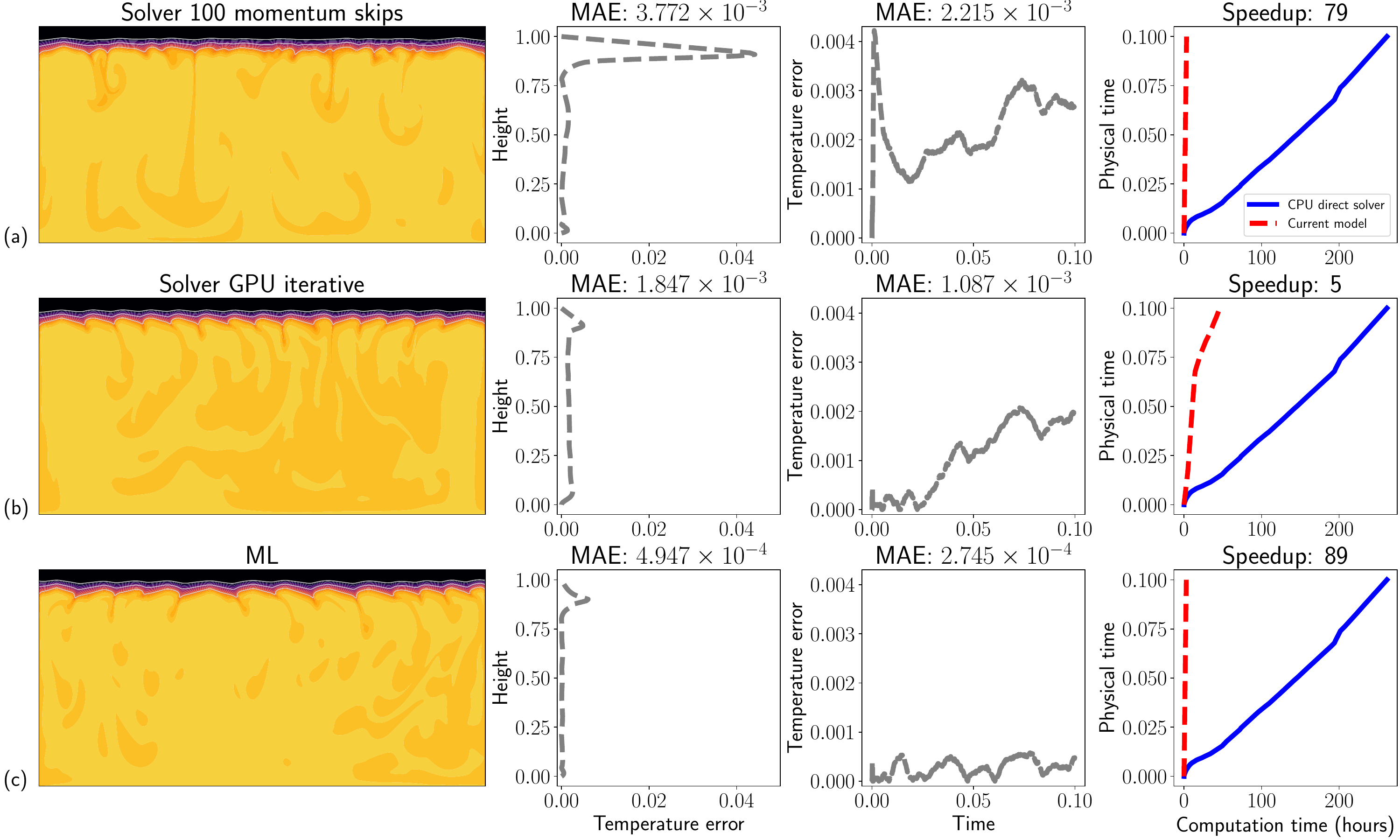}}
\caption{Comparison of accuracy and speedup of (a) direct solver with $100$ momentum skips on the CPU, (b) iterative solver with an under relaxation factor of $0.99$ on the GPU, and (c) our model with the ML-based Stokes solver running on the GPU.}
\label{fig-speedup}
\end{figure*}

To evaluate the speedup of the model, we select an unseen simulation from the test set and run four different models on the same CPU/GPU architecture using the same numerical settings. We run (1) the original direct solver on a CPU, (2) the direct solver with $100$ momentum skips on a CPU, (3) an iterative solver (BiCGStab) with under relaxation factor of $0.99$ on a GPU, and (4) our ML model on a GPU. Fig. \ref{fig-speedup}c shows that our model not only achieves the highest speedup of the four, but also has the lowest error in mean quantities with respect to the original solver. This is understandable for the direct solver with momentum skips, but rather unexpected for the iterative solver. Indeed, we see that the lid distribution for the iterative solver looks very realistic, but the error in the mean temperature seems to accumulate in this case. This is likely because we use an under relaxation factor of 0.99, which means that the velocities solved for are not as accurate as those predicted by the direct solver. As in Fig. \ref{fig-parameters}(a,b), we do not expect this increased accuracy of our method to hold up across the entire parameter range, but it further underscores the promise of machine learning for mantle convection simulations. 

We report the best speedups here when a single run is executed on our best GPU. In practice, however, this speedup is throttled by the exchange of data between the Python and C++ code as more runs are launched on a single machine in parallel. This is not ideal for larger campaigns, and we would like to address it in the future. One possibility would be to infer the ML model directly in C++ and achieve even greater speedups. Better GPUs than the V100 can further extend this advantage.

\subsection{Different ablations in the Stokes model}
\label{sec-results_ablations}

\begin{table*}
    \centering
    \begin{tabular}{ccccc}
     Relative improvement of learned padding $\left( \uparrow \right)$      
          \hspace{2cm}  &  Left &  Right &  Top & Bottom \\
    to zeros padding with 16 filters & 8.03 & 2.78 & 1.63 & 2.82 \\
    to zeros padding with 64 filters & 3.42 & 2.03 & 0.99 & 1.52 
    \end{tabular}
    \caption{We compute the MAE on the boundaries for every fifth sample in our entire test set for different paddings and plot them with respect to the MAE on the learned padding (boundary-learned convolution).}
    \label{tab:mae_pad}
\end{table*}

\begin{figure*}
\centerline{\includegraphics[width=\textwidth]{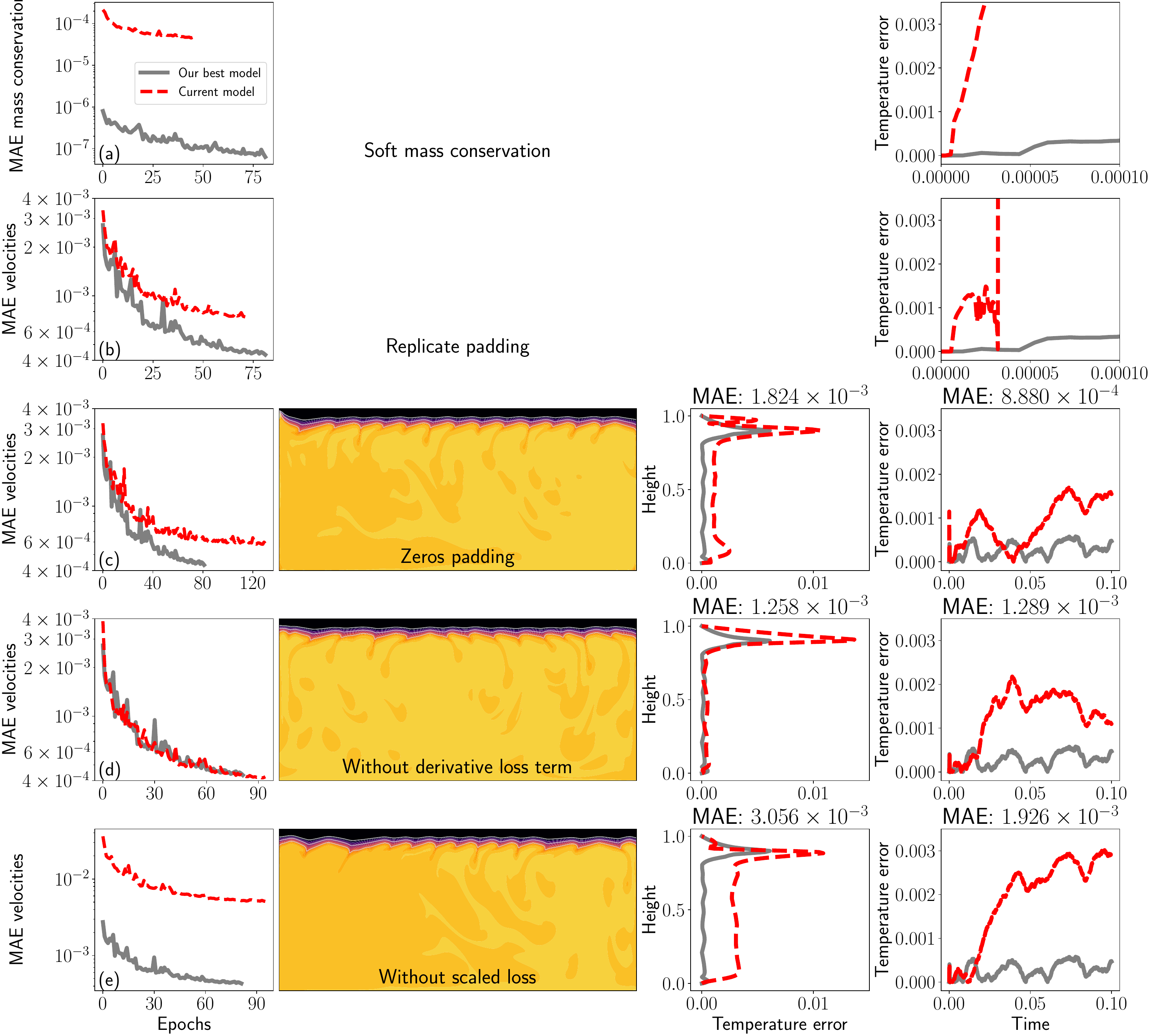}}
\caption{Results of different ablations, where we change one component of the model at a time. The cross-validation loss is plotted in the first column in solid grey for our best model and the ablated model in dashed red.}
\label{fig-ablations}
\end{figure*}

We now present the importance of specific components of our CNN through ablations, where we remove one component at a time while keeping the rest of the architecture unchanged. Fig. \ref{fig-ablations}a highlights the importance of enforcing mass conservation for producing a stable rollout. We add mass conservation as a soft constraint to the loss function in Eq. \eqref{eq:mass_loss}, and observe that the overall mass conservation curve in dashed red going down in Fig. \ref{fig-ablations}a. However, it is likely just a reflection of the velocity predictions getting better. In our case, where we have ground truth data available at every spatial point in the domain, the mass conservation is not much more than a transformation of data that does not seem to convey any more information during the training. Note, that when we enforce mass conservation as a hard constraint using the curl-based formulation, the loss on the interior domain is $0$ up to double machine precision. The grey solid curve in Fig. \ref{fig-ablations}b has higher values than double-precision zero because of the boundaries. Finding a way to include a curl-based constraint here could further improve accuracy. Nevertheless, the mass conservation error on boundaries decreases as the network predicts better and better velocities. We also tried to multiply the mass conservation loss term with a factor to give it more weight, but found that it conflicted with the other terms in that it deteriorated the loss values of velocities. This is a well-known issue with physics-based losses and could be addressed in future work \citep{liu2025config}.

Next, rollout performance was highly sensitive to errors on the boundaries. Zero padding outperforms replicate padding, but we still observe a severe artifact in Fig. \ref{fig-ablations}d near the upper-left corner of the domain, where the stagnant lid is ``pinched'' and some other smaller, more difficult to see inaccuracies appear at the bottom corners. Since the boundary learned convolution introduces a lot more trainable parameters, we also match the parameter count of the zero padding network by increasing the filter count from $16$ to $64$. Unfortunately, this deteriorated the performance even more as larger models can be more prone to diverge during rollout. We calculated the test set MAE on each boundary for the learned padding and compared it those of the zero padding networks. In Table \ref{tab:mae_pad}, we list these as relative improvements and notice that the learned padding is up to $8$ times more accurate than zero padding. 

Finally, we evaluate the importance of learning on derivatives of the velocities (Eq. \eqref{eq:loss_derivative}) and find modest improvements in the MAE of horizontally-averaged temperature profile and the mean temperature over time. When the loss is not scaled (Eq. \eqref{eq:loss_scaling}), the presence of the derivative term seems to create some imbalance in loss terms, but the deterioration in the MAE values stems mainly from the lack of loss scaling. When loss scaling and the derivative term are both taken out, the MAE is $2.804 \times 10^{-3}$ for the temperature profile and $1.294 \times 10^{-3}$ for the mean temperature vs. time curve. To summarize, both the loss tricks seem to help.

\newpage
\subsection{Towards a general Stokes model}
\label{sec-results_generalizability}

\begin{figure*}
\centerline{\includegraphics[width=\textwidth]{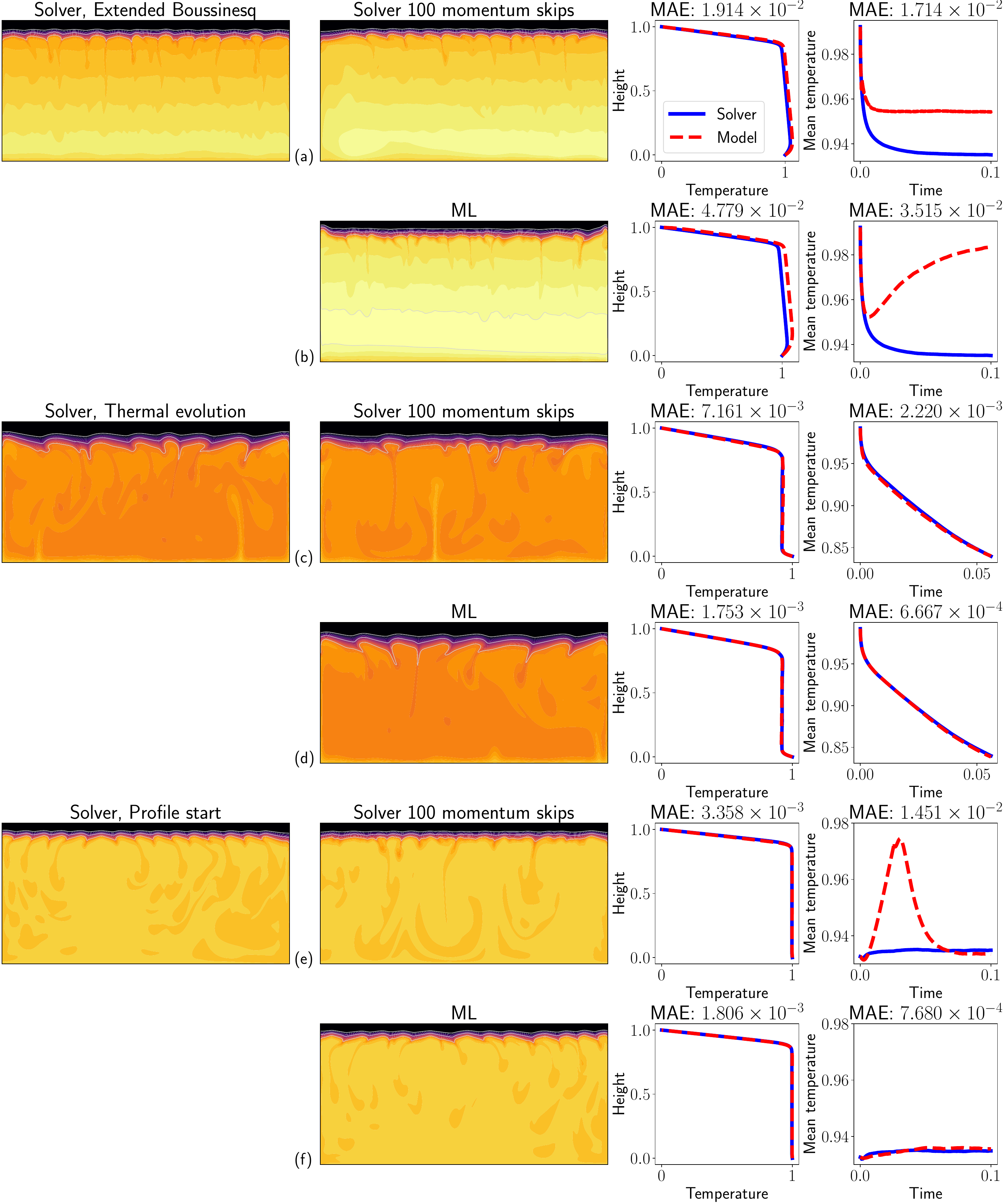}}
\caption{Performance of the baseline and the ML model on three unseen scenarios: (a,b) Extended Boussinesq Approximation (EBA), (c,d) thermal evolution, and (e-f) starting from the steady-state profile.}
\label{fig-ood}
\end{figure*}

With our physics-based machine learning approach, we are able to learn from a relatively small training dataset and generate high-quality predictions. We now test the robustness of this approach on the following unseen scenarios, some of which are out-of-distribution (OOD). To do so, we modify some aspect of the simulation presented in Fig. \ref{fig-parameters}e,f. 

\textbf{One}, we plug our Stokes surrogate into a different energy equation than the one training data was generated with. Specifically, we use the Extended Boussinesq Approximation (EBA), where the mass and momentum equations are unchanged, but the energy equation contains two extra terms (adiabatic heating and shear heating) associated with the degree of compressibility in the form of a Dissipation number ($Di$) \citep{king2010}. We run a simulation with a low $Di=0.1$, which leads to a small adiabatic temperature gradient between the top and bottom boundary layers and renders this an OOD example. As we see from Fig. \ref{fig-ood}a,b, our model fails to accurately match the ground truth, both in terms of mean quantities and the convection patterns including the noticeable thinning of the lid at the top corners. The solver with momentum skips fares slightly better. Nevertheless, we believe that including some data from EBA runs could alleviate this problem, possibly without introducing additional input variables such as $Di$ to our Stokes surrogate model.

\textbf{Two}, we run a thermal evolution simulation by introducing a time-decaying heat source. We run 
the simulation until a non-dimensional time $t=0.056$, which roughly corresponds to $4.5$ billion years for a Mars-like planet. A decay constant is set to reduce the heat production ($Q$ in Eq. \eqref{eq:energy}) by a factor of $10$ over the entire evolution. As a consequence, the simulation no longer enters a statistical steady-state but continues evolving with a decreasing mean temperature. Our model outperforms the baseline and produces a very realistic-looking convection pattern. This approach has some parallels to parametrized 1D thermal evolution models, which rely on scaling laws to predict convective heat fluxes that are then used to advance the mean temperature through a global energy balance equation \citep[e.g.,][]{stevenson1983,gurnis1989,solomatov1995,korenaga2003,Grott2011,drilleau2021}. In some respects, our approach can be seen as an extension of these methods in that it enables the prediction of the full spatio-temporal evolution of velocities and temperature from a purely steady-state dataset. 

\textbf{Three}, we try different initializations for the same simulation as also done in \citet{agarwal2025}. The model failed completely when initialized with a cold profile (zero temperature everywhere apart from the bottom boundary), as this lies far outside the distributions of profiles in our training set. The solver with momentum skips on the other hand still managed to follow the solution, although not perfectly. As opposed to the cold start, our model fared better when starting with a profile with a temperature of $0.75$ everywhere except a small linearly increasing thermal boundary layer of $0.05$ on top and bottom. Yet, upon visually inspecting an animation of the simulation, we found that our model does not mimic the solver in the conductive growth of the boundary layers in the initial phase. This is because the minimum internal temperature present in the training set is $0.8$, making this case slightly OOD. The fact that our model can still enter the statistical steady-state is interesting. We never teach the network to follow any trajectory or reach a certain end point. This behavior more likely hints at the inherent property of this type of flows: regardless of the initial transient stages, the flow will tend to reach the same statistical steady-state. It remains to be seen if this property can be exploited to discover steady-states even faster. Nevertheless, despite all the physics-based components in our model, it is a sobering reminder that ML models can struggle on OOD examples. In that respect, terminology such as neural solvers or ML-based simulations can be confusing as a certain degree of robustness is expected from solvers. As an example of a study that shows significant OOD capabilities, \citet{ranade2022a} take a data-driven approach for learning machine learning models on local domains in space, but then infer iteratively like a solver to propagate information globally and arrive at an equilibrium solution of the PDE. 

Finally, we use the profile prediction model from \citet{agarwal2025} to initialize our simulation and our prediction models and find that our model convincingly outperforms the baseline as can be seen in Fig.\ref{fig-ood}e,f. This is encouraging, as our model has never seen this type of initial condition in the training dataset, but the horizontally-averaged temperature field is, of course well within the distribution of the training set. Multiple models can thus be combined to further accelerate mantle convection simulations.

While it is essential to be aware of the limitations of these models, at the same time, these OOD cases provide a template for extending the training dataset. Training data from different initializations and the EBA energy equation can be seamlessly introduced into our existing model. It would also be interesting to extend the setup to different viscosity functions than in Eq. \eqref{eq:eta}. Doing so would require a different way of scaling the velocities -- we would like to be independent of the parameters $\beta$ and $\gamma$, as other formulations of viscosity include different parameters. For instance, one could instead scale the velocity based on the minimum value of the viscosity field. The velocity scaling would also need to be explored if more complex physics were introduced, e.g., in the form of pressure- and temperature-dependent thermal expansivity and thermal conductivity, phase transitions and melting. Higher-order optimizers \citep[e.g.,][]{shampoo,soap} could also be interesting for learning across several orders of magnitude. Lastly, in future work, we would like to explore discretization invariance and learning as a function of different geometries. Neural fields \citep{serrano2023operator, catalani_neural_2024, knigge2024spacetime} have shown promise in memory-efficient learning on arbitrary computational grids.

\section{Conclusion}

We present a novel physics-based machine learning approach for mantle convection simulations. We learn a single model to predict flow velocities spanning four orders of magnitude as a function of three parameters controlling viscosity and internal heating rate. These velocities are then fed to a numerical finite-volume solver to perform advection-diffusion of the temperature field and to obtain the next time-step. We found this approach promising for stable predictions over thousands of time-steps, an active area of research in machine learning for PDEs. On the contrary, we were unable to achieve stable rollout by learning in time, despite reaching lower validation losses on the velocities during training. Overall, our model provides a speedup of $89$ times, compared to the numerical solution computed with a direct solver.

Our model is stable across the parameter space, except when approaching high values of the internal heating rate ($Q$) and of the parameter controlling the temperature dependence of the viscosity ($\gamma$), as this region of the parameter space is not represented in our sparse training dataset of $94$ simulations. We also demonstrated the strengths of our model by removing one component at a time. Mass conservation plays a decisive role. Using boundary-learned convolutions, we obtain up to $8$-times more accurate solutions on the boundaries than in the case of zero padding, which further enhances our rollout performance. 

Finally, we tested our model on some unseen scenarios. When using a different energy equation including additional compressibility effects, we found that our model fails as the input temperature fields contain an adiabatic gradient in the convecting part, which is not considered in the simulations of our training set. We also tried to start our model from a cold profile, but found this was too far away from the training set. 

A particularly promising result is the ability to perform a thermal evolution simulation based on a purely steady-state dataset. This shows the versatility of our formulation, which is already able to generalize in some instances. Physics-based machine learning promises accurate and data-efficient modeling of mantle convection simulations.

\section*{Author Declarations}

\subsection*{Conflict of Interest}
\noindent The authors have no conflicts to disclose.

\subsection*{Author Contributions}

\noindent 
 \textit{Conceptualization}: SA, NT, DG, AB, CH; \\
\textit{Methodology}: SA, CH, AB, DG, NT; \\
\textit{Software}: SA, CH ; \\
\textit{Validation}: SA  ; \\
\textit{Formal analysis}: SA ; \\
\textit{Investigation}: SA  ; \\
\textit{Resources}: NT, CH  ; \\
\textit{Data curation}: NT, SA ; \\
\textit{Writing–Original Draft}: SA, AB, NT; \\
\textit{Writing–Review \& Editing}: NT, DG, CH, AB, SA ; \\
\textit{Visualization}: SA; \\
\textit{Supervision}: NT, DG;  \\
\textit{Project administration}: NT, DG; \\
\textit{Funding acquisition}: NT, DG, SA; \\

\begin{acknowledgments}
This work was funded by the PLAGeS (Physics-based Learning Algorithms for Geophysical flow Simulations) project through the the German  Ministry of Education and Research BMBF (project numbers 16DKWN117A and 16DKWN117B) under the ``Deutschen Aufbau- und Resilienzplans'' (DARP). DARP is part of NextGenerationEU's ``Aufbau- und Resilienzfazilität'' (ARF) by the European Union.
\end{acknowledgments}

\section*{Data Availability}

Some sample data from the simulations as well as model weights for the trained networks are available in \citet{agarwal_2024b_zenodo}. These can be used to predict velocities here \url{https://github.com/agsiddhant/PBML_Mantle_Convection}. The Github repository contains all the machine learning code used in this paper. The simulations were generated at the German Aerospace Center (DLR). The complete derived data supporting the findings of this study are available from the corresponding author upon reasonable request. Furthermore, GAIA is a proprietary code of the DLR and users interested in working with it should contact Nicola Tosi (nicola.tosi@dlr.de) and Christian H{\"u}ttig (christian.huettig@dlr.de).

\bibliography{Bibliography}

\begin{thebibliography}{58}%
\makeatletter
\providecommand \@ifxundefined [1]{%
 \@ifx{#1\undefined}
}%
\providecommand \@ifnum [1]{%
 \ifnum #1\expandafter \@firstoftwo
 \else \expandafter \@secondoftwo
 \fi
}%
\providecommand \@ifx [1]{%
 \ifx #1\expandafter \@firstoftwo
 \else \expandafter \@secondoftwo
 \fi
}%
\providecommand \natexlab [1]{#1}%
\providecommand \enquote  [1]{``#1''}%
\providecommand \bibnamefont  [1]{#1}%
\providecommand \bibfnamefont [1]{#1}%
\providecommand \citenamefont [1]{#1}%
\providecommand \href@noop [0]{\@secondoftwo}%
\providecommand \href [0]{\begingroup \@sanitize@url \@href}%
\providecommand \@href[1]{\@@startlink{#1}\@@href}%
\providecommand \@@href[1]{\endgroup#1\@@endlink}%
\providecommand \@sanitize@url [0]{\catcode `\\12\catcode `\$12\catcode `\&12\catcode `\#12\catcode `\^12\catcode `\_12\catcode `\%12\relax}%
\providecommand \@@startlink[1]{}%
\providecommand \@@endlink[0]{}%
\providecommand \url  [0]{\begingroup\@sanitize@url \@url }%
\providecommand \@url [1]{\endgroup\@href {#1}{\urlprefix }}%
\providecommand \urlprefix  [0]{URL }%
\providecommand \Eprint [0]{\href }%
\providecommand \doibase [0]{http://dx.doi.org/}%
\providecommand \selectlanguage [0]{\@gobble}%
\providecommand \bibinfo  [0]{\@secondoftwo}%
\providecommand \bibfield  [0]{\@secondoftwo}%
\providecommand \translation [1]{[#1]}%
\providecommand \BibitemOpen [0]{}%
\providecommand \bibitemStop [0]{}%
\providecommand \bibitemNoStop [0]{.\EOS\space}%
\providecommand \EOS [0]{\spacefactor3000\relax}%
\providecommand \BibitemShut  [1]{\csname bibitem#1\endcsname}%
\let\auto@bib@innerbib\@empty
\bibitem [{\citenamefont {Agarwal}\ \emph {et~al.}(2020)\citenamefont {Agarwal}, \citenamefont {Tosi}, \citenamefont {Breuer}, \citenamefont {Padovan}, \citenamefont {Kessel},\ and\ \citenamefont {Montavon}}]{agarwal2020}%
  \BibitemOpen
  \bibfield  {author} {\bibinfo {author} {\bibnamefont {Agarwal}, \bibfnamefont {S.}}, \bibinfo {author} {\bibnamefont {Tosi}, \bibfnamefont {N.}}, \bibinfo {author} {\bibnamefont {Breuer}, \bibfnamefont {D.}}, \bibinfo {author} {\bibnamefont {Padovan}, \bibfnamefont {S.}}, \bibinfo {author} {\bibnamefont {Kessel}, \bibfnamefont {P.}}, \ and\ \bibinfo {author} {\bibnamefont {Montavon}, \bibfnamefont {G.}},\ }\bibfield  {title} {\enquote {\bibinfo {title} {{A machine-learning-based surrogate model of Mars’ thermal evolution}},}\ }\href {\doibase 10.1093/gji/ggaa234} {\bibfield  {journal} {\bibinfo  {journal} {Geophysical Journal International}\ }\textbf {\bibinfo {volume} {222}},\ \bibinfo {pages} {1656--1670} (\bibinfo {year} {2020})}\BibitemShut {NoStop}%
\bibitem [{\citenamefont {Agarwal}\ \emph {et~al.}(2025)\citenamefont {Agarwal}, \citenamefont {Tosi}, \citenamefont {Hüttig}, \citenamefont {Greenberg},\ and\ \citenamefont {Bekar}}]{agarwal_2024b_zenodo}%
  \BibitemOpen
  \bibfield  {author} {\bibinfo {author} {\bibnamefont {Agarwal}, \bibfnamefont {S.}}, \bibinfo {author} {\bibnamefont {Tosi}, \bibfnamefont {N.}}, \bibinfo {author} {\bibnamefont {Hüttig}, \bibfnamefont {C.}}, \bibinfo {author} {\bibnamefont {Greenberg}, \bibfnamefont {D.}}, \ and\ \bibinfo {author} {\bibnamefont {Bekar}, \bibfnamefont {A.}},\ }\href {\doibase 10.5281/zenodo.15088589} {\enquote {\bibinfo {title} {{PBML\_Mantle\_Convection [Dataset]}},}\ } (\bibinfo {year} {2025})\BibitemShut {NoStop}%
\bibitem [{\citenamefont {Agarwal}\ \emph {et~al.}(2024)\citenamefont {Agarwal}, \citenamefont {Tosi}, \citenamefont {Hüttig}, \citenamefont {Greenberg},\ and\ \citenamefont {Bekar}}]{agarwal2025}%
  \BibitemOpen
  \bibfield  {author} {\bibinfo {author} {\bibnamefont {Agarwal}, \bibfnamefont {S.}}, \bibinfo {author} {\bibnamefont {Tosi}, \bibfnamefont {N.}}, \bibinfo {author} {\bibnamefont {Hüttig}, \bibfnamefont {C.}}, \bibinfo {author} {\bibnamefont {Greenberg}, \bibfnamefont {D.~S.}}, \ and\ \bibinfo {author} {\bibnamefont {Bekar}, \bibfnamefont {A.~C.}},\ }\href {https://arxiv.org/abs/2408.17298} {\enquote {\bibinfo {title} {Accelerating the discovery of steady-states of planetary interior dynamics with machine learning},}\ } (\bibinfo {year} {2024}),\ \Eprint {http://arxiv.org/abs/2408.17298} {arXiv:2408.17298 [physics.flu-dyn]} \BibitemShut {NoStop}%
\bibitem [{\citenamefont {Agarwal}\ \emph {et~al.}(2021)\citenamefont {Agarwal}, \citenamefont {Tosi}, \citenamefont {Kessel}, \citenamefont {Breuer},\ and\ \citenamefont {Montavon}}]{agarwal2021b}%
  \BibitemOpen
  \bibfield  {author} {\bibinfo {author} {\bibnamefont {Agarwal}, \bibfnamefont {S.}}, \bibinfo {author} {\bibnamefont {Tosi}, \bibfnamefont {N.}}, \bibinfo {author} {\bibnamefont {Kessel}, \bibfnamefont {P.}}, \bibinfo {author} {\bibnamefont {Breuer}, \bibfnamefont {D.}}, \ and\ \bibinfo {author} {\bibnamefont {Montavon}, \bibfnamefont {G.}},\ }\bibfield  {title} {\enquote {\bibinfo {title} {Deep learning for surrogate modeling of two-dimensional mantle convection},}\ }\href {\doibase 10.1103/PhysRevFluids.6.113801} {\bibfield  {journal} {\bibinfo  {journal} {Phys. Rev. Fluids}\ }\textbf {\bibinfo {volume} {6}},\ \bibinfo {pages} {113801} (\bibinfo {year} {2021})}\BibitemShut {NoStop}%
\bibitem [{\citenamefont {Alguacil}\ \emph {et~al.}(2021)\citenamefont {Alguacil}, \citenamefont {Pinto}, \citenamefont {Bauerheim}, \citenamefont {Jacob},\ and\ \citenamefont {Moreau}}]{cnnpadding}%
  \BibitemOpen
  \bibfield  {author} {\bibinfo {author} {\bibnamefont {Alguacil}, \bibfnamefont {A.}}, \bibinfo {author} {\bibnamefont {Pinto}, \bibfnamefont {W.~G.}}, \bibinfo {author} {\bibnamefont {Bauerheim}, \bibfnamefont {M.}}, \bibinfo {author} {\bibnamefont {Jacob}, \bibfnamefont {M.~C.}}, \ and\ \bibinfo {author} {\bibnamefont {Moreau}, \bibfnamefont {S.}},\ }\bibfield  {title} {\enquote {\bibinfo {title} {Effects of boundary conditions in fully convolutional networks for learning spatio-temporal dynamics},}\ }in\ \href@noop {} {\emph {\bibinfo {booktitle} {Machine Learning and Knowledge Discovery in Databases. Applied Data Science Track}}},\ \bibinfo {editor} {edited by\ \bibinfo {editor} {\bibfnamefont {Y.}~\bibnamefont {Dong}}, \bibinfo {editor} {\bibfnamefont {N.}~\bibnamefont {Kourtellis}}, \bibinfo {editor} {\bibfnamefont {B.}~\bibnamefont {Hammer}}, \ and\ \bibinfo {editor} {\bibfnamefont {J.~A.}\ \bibnamefont {Lozano}}}\ (\bibinfo  {publisher} {Springer International Publishing},\ \bibinfo {address}
  {Cham},\ \bibinfo {year} {2021})\ pp.\ \bibinfo {pages} {102--117}\BibitemShut {NoStop}%
\bibitem [{\citenamefont {Alieva}\ \emph {et~al.}(2023)\citenamefont {Alieva}, \citenamefont {Hoyer}, \citenamefont {Brenner}, \citenamefont {Iaccarino},\ and\ \citenamefont {Norgaard}}]{alieva_2023}%
  \BibitemOpen
  \bibfield  {author} {\bibinfo {author} {\bibnamefont {Alieva}, \bibfnamefont {A.}}, \bibinfo {author} {\bibnamefont {Hoyer}, \bibfnamefont {S.}}, \bibinfo {author} {\bibnamefont {Brenner}, \bibfnamefont {M.}}, \bibinfo {author} {\bibnamefont {Iaccarino}, \bibfnamefont {G.}}, \ and\ \bibinfo {author} {\bibnamefont {Norgaard}, \bibfnamefont {P.}},\ }\bibfield  {title} {\enquote {\bibinfo {title} {Toward accelerated data-driven rayleigh--b{\'e}nard convection simulations},}\ }\href {\doibase 10.1140/epje/s10189-023-00302-w} {\bibfield  {journal} {\bibinfo  {journal} {The European Physical Journal E}\ }\textbf {\bibinfo {volume} {46}},\ \bibinfo {pages} {64} (\bibinfo {year} {2023})}\BibitemShut {NoStop}%
\bibitem [{\citenamefont {Amestoy}\ \emph {et~al.}(2019)\citenamefont {Amestoy}, \citenamefont {Buttari}, \citenamefont {L'Excellent},\ and\ \citenamefont {Mary}}]{MUMPS:2}%
  \BibitemOpen
  \bibfield  {author} {\bibinfo {author} {\bibnamefont {Amestoy}, \bibfnamefont {P.}}, \bibinfo {author} {\bibnamefont {Buttari}, \bibfnamefont {A.}}, \bibinfo {author} {\bibnamefont {L'Excellent}, \bibfnamefont {J.-Y.}}, \ and\ \bibinfo {author} {\bibnamefont {Mary}, \bibfnamefont {T.}},\ }\bibfield  {title} {\enquote {\bibinfo {title} {{Performance and Scalability of the Block Low-Rank Multifrontal Factorization on Multicore Architectures}},}\ }\href@noop {} {\bibfield  {journal} {\bibinfo  {journal} {ACM Transactions on Mathematical Software}\ }\textbf {\bibinfo {volume} {45}},\ \bibinfo {pages} {2:1--2:26} (\bibinfo {year} {2019})}\BibitemShut {NoStop}%
\bibitem [{\citenamefont {Amestoy}\ \emph {et~al.}(2001)\citenamefont {Amestoy}, \citenamefont {Duff}, \citenamefont {Koster},\ and\ \citenamefont {L'Excellent}}]{MUMPS:1}%
  \BibitemOpen
  \bibfield  {author} {\bibinfo {author} {\bibnamefont {Amestoy}, \bibfnamefont {P.}}, \bibinfo {author} {\bibnamefont {Duff}, \bibfnamefont {I.~S.}}, \bibinfo {author} {\bibnamefont {Koster}, \bibfnamefont {J.}}, \ and\ \bibinfo {author} {\bibnamefont {L'Excellent}, \bibfnamefont {J.-Y.}},\ }\bibfield  {title} {\enquote {\bibinfo {title} {A fully asynchronous multifrontal solver using distributed dynamic scheduling},}\ }\href@noop {} {\bibfield  {journal} {\bibinfo  {journal} {SIAM Journal on Matrix Analysis and Applications}\ }\textbf {\bibinfo {volume} {23}},\ \bibinfo {pages} {15--41} (\bibinfo {year} {2001})}\BibitemShut {NoStop}%
\bibitem [{\citenamefont {Bangerth}\ \emph {et~al.}(2024)\citenamefont {Bangerth}, \citenamefont {Dannberg}, \citenamefont {Fraters}, \citenamefont {Gassmoeller}, \citenamefont {Glerum}, \citenamefont {Heister}, \citenamefont {Myhill},\ and\ \citenamefont {Naliboff}}]{aspect-doi-v3.0.0}%
  \BibitemOpen
  \bibfield  {author} {\bibinfo {author} {\bibnamefont {Bangerth}, \bibfnamefont {W.}}, \bibinfo {author} {\bibnamefont {Dannberg}, \bibfnamefont {J.}}, \bibinfo {author} {\bibnamefont {Fraters}, \bibfnamefont {M.}}, \bibinfo {author} {\bibnamefont {Gassmoeller}, \bibfnamefont {R.}}, \bibinfo {author} {\bibnamefont {Glerum}, \bibfnamefont {A.}}, \bibinfo {author} {\bibnamefont {Heister}, \bibfnamefont {T.}}, \bibinfo {author} {\bibnamefont {Myhill}, \bibfnamefont {R.}}, \ and\ \bibinfo {author} {\bibnamefont {Naliboff}, \bibfnamefont {J.}},\ }\href {\doibase 10.5281/zenodo.14371679} {\enquote {\bibinfo {title} {Aspect v3.0.0},}\ } (\bibinfo {year} {2024})\BibitemShut {NoStop}%
\bibitem [{\citenamefont {Breuer}\ and\ \citenamefont {Moore}(2015)}]{breuer2015}%
  \BibitemOpen
  \bibfield  {author} {\bibinfo {author} {\bibnamefont {Breuer}, \bibfnamefont {D.}}\ and\ \bibinfo {author} {\bibnamefont {Moore}, \bibfnamefont {W.}},\ }\bibfield  {title} {\enquote {\bibinfo {title} {Dynamics and thermal history of the terrestrial planets, the moon, and io},}\ }in\ \href {\doibase 10.1016/B978-0-444-53802-4.00173-1} {\emph {\bibinfo {booktitle} {Treatise on Geophysics (Second Edition)}}},\ Vol.~\bibinfo {volume} {10},\ \bibinfo {editor} {edited by\ \bibinfo {editor} {\bibfnamefont {G.}~\bibnamefont {Schubert}}}\ (\bibinfo  {publisher} {Elsevier},\ \bibinfo {address} {Oxford},\ \bibinfo {year} {2015})\ \bibinfo {edition} {2nd}\ ed.,\ pp.\ \bibinfo {pages} {255 -- 305}\BibitemShut {NoStop}%
\bibitem [{\citenamefont {Buitrago}\ \emph {et~al.}(2025)\citenamefont {Buitrago}, \citenamefont {Marwah}, \citenamefont {Gu},\ and\ \citenamefont {Risteski}}]{buitrago2025on}%
  \BibitemOpen
  \bibfield  {author} {\bibinfo {author} {\bibnamefont {Buitrago}, \bibfnamefont {R.}}, \bibinfo {author} {\bibnamefont {Marwah}, \bibfnamefont {T.}}, \bibinfo {author} {\bibnamefont {Gu}, \bibfnamefont {A.}}, \ and\ \bibinfo {author} {\bibnamefont {Risteski}, \bibfnamefont {A.}},\ }\bibfield  {title} {\enquote {\bibinfo {title} {On the benefits of memory for modeling time-dependent {PDE}s},}\ }in\ \href {https://openreview.net/forum?id=o9kqa5K3tB} {\emph {\bibinfo {booktitle} {The Thirteenth International Conference on Learning Representations}}}\ (\bibinfo {year} {2025})\BibitemShut {NoStop}%
\bibitem [{\citenamefont {Catalani}\ \emph {et~al.}(2024)\citenamefont {Catalani}, \citenamefont {Agarwal}, \citenamefont {Bertrand}, \citenamefont {Tost}, \citenamefont {Bauerheim},\ and\ \citenamefont {Morlier}}]{catalani_neural_2024}%
  \BibitemOpen
  \bibfield  {author} {\bibinfo {author} {\bibnamefont {Catalani}, \bibfnamefont {G.}}, \bibinfo {author} {\bibnamefont {Agarwal}, \bibfnamefont {S.}}, \bibinfo {author} {\bibnamefont {Bertrand}, \bibfnamefont {X.}}, \bibinfo {author} {\bibnamefont {Tost}, \bibfnamefont {F.}}, \bibinfo {author} {\bibnamefont {Bauerheim}, \bibfnamefont {M.}}, \ and\ \bibinfo {author} {\bibnamefont {Morlier}, \bibfnamefont {J.}},\ }\bibfield  {title} {\enquote {\bibinfo {title} {Neural fields for rapid aircraft aerodynamics simulations},}\ }\href {\doibase 10.1038/s41598-024-76983-w} {\bibfield  {journal} {\bibinfo  {journal} {Scientific Reports}\ }\textbf {\bibinfo {volume} {14}},\ \bibinfo {pages} {25496} (\bibinfo {year} {2024})},\ \bibinfo {note} {publisher: Nature Publishing Group}\BibitemShut {NoStop}%
\bibitem [{\citenamefont {Cuomo}\ \emph {et~al.}(2022)\citenamefont {Cuomo}, \citenamefont {Di~Cola}, \citenamefont {Giampaolo}, \citenamefont {Rozza}, \citenamefont {Raissi},\ and\ \citenamefont {Piccialli}}]{cuomo_scientific_2022}%
  \BibitemOpen
  \bibfield  {author} {\bibinfo {author} {\bibnamefont {Cuomo}, \bibfnamefont {S.}}, \bibinfo {author} {\bibnamefont {Di~Cola}, \bibfnamefont {V.~S.}}, \bibinfo {author} {\bibnamefont {Giampaolo}, \bibfnamefont {F.}}, \bibinfo {author} {\bibnamefont {Rozza}, \bibfnamefont {G.}}, \bibinfo {author} {\bibnamefont {Raissi}, \bibfnamefont {M.}}, \ and\ \bibinfo {author} {\bibnamefont {Piccialli}, \bibfnamefont {F.}},\ }\bibfield  {title} {\enquote {\bibinfo {title} {Scientific {Machine} {Learning} {Through} {Physics}–{Informed} {Neural} {Networks}: {Where} we are and {What}’s {Next}},}\ }\href {\doibase 10.1007/s10915-022-01939-z} {\bibfield  {journal} {\bibinfo  {journal} {Journal of Scientific Computing}\ }\textbf {\bibinfo {volume} {92}},\ \bibinfo {pages} {88} (\bibinfo {year} {2022})}\BibitemShut {NoStop}%
\bibitem [{\citenamefont {Deschamps}\ and\ \citenamefont {Sotin}(2001)}]{deschamps2001}%
  \BibitemOpen
  \bibfield  {author} {\bibinfo {author} {\bibnamefont {Deschamps}, \bibfnamefont {F.}}\ and\ \bibinfo {author} {\bibnamefont {Sotin}, \bibfnamefont {C.}},\ }\bibfield  {title} {\enquote {\bibinfo {title} {Thermal convection in the outer shell of large icy satellites},}\ }\href@noop {} {\bibfield  {journal} {\bibinfo  {journal} {Journal of Geophysical Research - Planets}\ }\textbf {\bibinfo {volume} {106}},\ \bibinfo {pages} {5107--5121} (\bibinfo {year} {2001})}\BibitemShut {NoStop}%
\bibitem [{\citenamefont {Drilleau}\ \emph {et~al.}(2021)\citenamefont {Drilleau}, \citenamefont {Samuel}, \citenamefont {Rivoldini}, \citenamefont {Panning},\ and\ \citenamefont {Lognonné}}]{drilleau2021}%
  \BibitemOpen
  \bibfield  {author} {\bibinfo {author} {\bibnamefont {Drilleau}, \bibfnamefont {M.}}, \bibinfo {author} {\bibnamefont {Samuel}, \bibfnamefont {H.}}, \bibinfo {author} {\bibnamefont {Rivoldini}, \bibfnamefont {A.}}, \bibinfo {author} {\bibnamefont {Panning}, \bibfnamefont {M.}}, \ and\ \bibinfo {author} {\bibnamefont {Lognonné}, \bibfnamefont {P.}},\ }\bibfield  {title} {\enquote {\bibinfo {title} {{Bayesian inversion of the Martian structure using geodynamic constraints}},}\ }\href {\doibase 10.1093/gji/ggab105} {\bibfield  {journal} {\bibinfo  {journal} {Geophysical Journal International}\ }\textbf {\bibinfo {volume} {226}},\ \bibinfo {pages} {1615--1644} (\bibinfo {year} {2021})}\BibitemShut {NoStop}%
\bibitem [{\citenamefont {Dumoulin}, \citenamefont {Doin},\ and\ \citenamefont {Fleitout}(1999)}]{dumoulin1999}%
  \BibitemOpen
  \bibfield  {author} {\bibinfo {author} {\bibnamefont {Dumoulin}, \bibfnamefont {C.}}, \bibinfo {author} {\bibnamefont {Doin}, \bibfnamefont {M.-P.}}, \ and\ \bibinfo {author} {\bibnamefont {Fleitout}, \bibfnamefont {L.}},\ }\bibfield  {title} {\enquote {\bibinfo {title} {Heat transport in stagnant lid convection with temperature- and pressure-dependent newtonian or non-newtonian rheology},}\ }\href {\doibase 10.1029/1999JB900110} {\bibfield  {journal} {\bibinfo  {journal} {Journal of Geophysical Research: Solid Earth}\ }\textbf {\bibinfo {volume} {104}},\ \bibinfo {pages} {12759--12777} (\bibinfo {year} {1999})}\BibitemShut {NoStop}%
\bibitem [{\citenamefont {Grimm}, \citenamefont {Heinlein},\ and\ \citenamefont {Klawonn}(2022)}]{kups64227}%
  \BibitemOpen
  \bibfield  {author} {\bibinfo {author} {\bibnamefont {Grimm}, \bibfnamefont {V.}}, \bibinfo {author} {\bibnamefont {Heinlein}, \bibfnamefont {A.}}, \ and\ \bibinfo {author} {\bibnamefont {Klawonn}, \bibfnamefont {A.}},\ }\href {https://kups.ub.uni-koeln.de/64227/} {\enquote {\bibinfo {title} {A short note on solving partial differential equations using convolutional neural networks},}\ }\bibinfo {type} {Technical Report}\ (\bibinfo  {institution} {Universit{\"a}t zu K{\"o}ln},\ \bibinfo {year} {2022})\BibitemShut {NoStop}%
\bibitem [{\citenamefont {Grott}, \citenamefont {Breuer},\ and\ \citenamefont {Laneuville}(2011)}]{Grott2011}%
  \BibitemOpen
  \bibfield  {author} {\bibinfo {author} {\bibnamefont {Grott}, \bibfnamefont {M.}}, \bibinfo {author} {\bibnamefont {Breuer}, \bibfnamefont {D.}}, \ and\ \bibinfo {author} {\bibnamefont {Laneuville}, \bibfnamefont {M.}},\ }\bibfield  {title} {\enquote {\bibinfo {title} {Thermo-chemical evolution and global contraction of mercury},}\ }\href {\doibase 10.1016/j.epsl.2011.04.040} {\bibfield  {journal} {\bibinfo  {journal} {Earth and Planetary Science Letters}\ }\textbf {\bibinfo {volume} {307}},\ \bibinfo {pages} {135 -- 146} (\bibinfo {year} {2011})}\BibitemShut {NoStop}%
\bibitem [{\citenamefont {Gupta}, \citenamefont {Koren},\ and\ \citenamefont {Singer}(2018)}]{shampoo}%
  \BibitemOpen
  \bibfield  {author} {\bibinfo {author} {\bibnamefont {Gupta}, \bibfnamefont {V.}}, \bibinfo {author} {\bibnamefont {Koren}, \bibfnamefont {T.}}, \ and\ \bibinfo {author} {\bibnamefont {Singer}, \bibfnamefont {Y.}},\ }\bibfield  {title} {\enquote {\bibinfo {title} {Shampoo: Preconditioned stochastic tensor optimization},}\ }in\ \href@noop {} {\emph {\bibinfo {booktitle} {International Conference on Machine Learning}}}\ (\bibinfo {year} {2018})\BibitemShut {NoStop}%
\bibitem [{\citenamefont {{Gurnis}}(1989)}]{gurnis1989}%
  \BibitemOpen
  \bibfield  {author} {\bibinfo {author} {\bibnamefont {{Gurnis}}, \bibfnamefont {M.}},\ }\bibfield  {title} {\enquote {\bibinfo {title} {{A reassessment of the heat transport by variable viscosity convection with plates and lids}},}\ }\href {\doibase 10.1029/GL016i002p00179} {\bibfield  {journal} {\bibinfo  {journal} {Geophy. Res. Lett.}\ }\textbf {\bibinfo {volume} {16}},\ \bibinfo {pages} {179--182} (\bibinfo {year} {1989})}\BibitemShut {NoStop}%
\bibitem [{\citenamefont {H\"uttig}, \citenamefont {Tosi},\ and\ \citenamefont {Moore}(2013)}]{huettig2013}%
  \BibitemOpen
  \bibfield  {author} {\bibinfo {author} {\bibnamefont {H\"uttig}, \bibfnamefont {C.}}, \bibinfo {author} {\bibnamefont {Tosi}, \bibfnamefont {N.}}, \ and\ \bibinfo {author} {\bibnamefont {Moore}, \bibfnamefont {W.}},\ }\bibfield  {title} {\enquote {\bibinfo {title} {An improved formulation of the incompressible navier-stokes equations with variable viscosity},}\ }\href {\doibase 10.1016/j.pepi.2013.04.002} {\bibfield  {journal} {\bibinfo  {journal} {Physics of the Earth and Planetary Interiors}\ }\textbf {\bibinfo {volume} {220}},\ \bibinfo {pages} {11--18} (\bibinfo {year} {2013})}\BibitemShut {NoStop}%
\bibitem [{\citenamefont {Innamorati}\ \emph {et~al.}(2020)\citenamefont {Innamorati}, \citenamefont {Ritschel}, \citenamefont {Weyrich},\ and\ \citenamefont {Mitra}}]{innamorati_learning_2020}%
  \BibitemOpen
  \bibfield  {author} {\bibinfo {author} {\bibnamefont {Innamorati}, \bibfnamefont {C.}}, \bibinfo {author} {\bibnamefont {Ritschel}, \bibfnamefont {T.}}, \bibinfo {author} {\bibnamefont {Weyrich}, \bibfnamefont {T.}}, \ and\ \bibinfo {author} {\bibnamefont {Mitra}, \bibfnamefont {N.~J.}},\ }\bibfield  {title} {\enquote {\bibinfo {title} {Learning on the {Edge}: {Investigating} {Boundary} {Filters} in {CNNs}},}\ }\href {\doibase 10.1007/s11263-019-01223-y} {\bibfield  {journal} {\bibinfo  {journal} {International Journal of Computer Vision}\ }\textbf {\bibinfo {volume} {128}},\ \bibinfo {pages} {773--782} (\bibinfo {year} {2020})}\BibitemShut {NoStop}%
\bibitem [{\citenamefont {{King}}\ \emph {et~al.}(2010)\citenamefont {{King}}, \citenamefont {{Lee}}, \citenamefont {{van Keken}}, \citenamefont {{Leng}}, \citenamefont {{Zhong}}, \citenamefont {{Tan}}, \citenamefont {{Tosi}},\ and\ \citenamefont {{Kameyama}}}]{king2010}%
  \BibitemOpen
  \bibfield  {author} {\bibinfo {author} {\bibnamefont {{King}}, \bibfnamefont {S.~D.}}, \bibinfo {author} {\bibnamefont {{Lee}}, \bibfnamefont {C.}}, \bibinfo {author} {\bibnamefont {{van Keken}}, \bibfnamefont {P.~E.}}, \bibinfo {author} {\bibnamefont {{Leng}}, \bibfnamefont {W.}}, \bibinfo {author} {\bibnamefont {{Zhong}}, \bibfnamefont {S.}}, \bibinfo {author} {\bibnamefont {{Tan}}, \bibfnamefont {E.}}, \bibinfo {author} {\bibnamefont {{Tosi}}, \bibfnamefont {N.}}, \ and\ \bibinfo {author} {\bibnamefont {{Kameyama}}, \bibfnamefont {M.~C.}},\ }\bibfield  {title} {\enquote {\bibinfo {title} {{A community benchmark for 2-D Cartesian compressible convection in the Earth's mantle}},}\ }\href {\doibase 10.1111/j.1365-246X.2009.04413.x} {\bibfield  {journal} {\bibinfo  {journal} {Geophysical Journal International}\ }\textbf {\bibinfo {volume} {180}},\ \bibinfo {pages} {73--87} (\bibinfo {year} {2010})}\BibitemShut {NoStop}%
\bibitem [{\citenamefont {Knigge}\ \emph {et~al.}(2024)\citenamefont {Knigge}, \citenamefont {Wessels}, \citenamefont {Valperga}, \citenamefont {Papa}, \citenamefont {Sonke}, \citenamefont {Bekkers},\ and\ \citenamefont {Gavves}}]{knigge2024spacetime}%
  \BibitemOpen
  \bibfield  {author} {\bibinfo {author} {\bibnamefont {Knigge}, \bibfnamefont {D.~M.}}, \bibinfo {author} {\bibnamefont {Wessels}, \bibfnamefont {D.}}, \bibinfo {author} {\bibnamefont {Valperga}, \bibfnamefont {R.}}, \bibinfo {author} {\bibnamefont {Papa}, \bibfnamefont {S.}}, \bibinfo {author} {\bibnamefont {Sonke}, \bibfnamefont {J.-J.}}, \bibinfo {author} {\bibnamefont {Bekkers}, \bibfnamefont {E.~J.}}, \ and\ \bibinfo {author} {\bibnamefont {Gavves}, \bibfnamefont {S.}},\ }\bibfield  {title} {\enquote {\bibinfo {title} {Space-time continuous {PDE} forecasting using equivariant neural fields},}\ }in\ \href {https://openreview.net/forum?id=wN5AgP0DJ0} {\emph {\bibinfo {booktitle} {The Thirty-eighth Annual Conference on Neural Information Processing Systems}}}\ (\bibinfo {year} {2024})\BibitemShut {NoStop}%
\bibitem [{\citenamefont {Kochkov}\ \emph {et~al.}(2021)\citenamefont {Kochkov}, \citenamefont {Smith}, \citenamefont {Alieva}, \citenamefont {Wang}, \citenamefont {Brenner},\ and\ \citenamefont {Hoyer}}]{Kochkov}%
  \BibitemOpen
  \bibfield  {author} {\bibinfo {author} {\bibnamefont {Kochkov}, \bibfnamefont {D.}}, \bibinfo {author} {\bibnamefont {Smith}, \bibfnamefont {J.~A.}}, \bibinfo {author} {\bibnamefont {Alieva}, \bibfnamefont {A.}}, \bibinfo {author} {\bibnamefont {Wang}, \bibfnamefont {Q.}}, \bibinfo {author} {\bibnamefont {Brenner}, \bibfnamefont {M.~P.}}, \ and\ \bibinfo {author} {\bibnamefont {Hoyer}, \bibfnamefont {S.}},\ }\bibfield  {title} {\enquote {\bibinfo {title} {Machine learning–accelerated computational fluid dynamics},}\ }\href {\doibase 10.1073/pnas.2101784118} {\bibfield  {journal} {\bibinfo  {journal} {Proceedings of the National Academy of Sciences}\ }\textbf {\bibinfo {volume} {118}},\ \bibinfo {pages} {e2101784118} (\bibinfo {year} {2021})}\BibitemShut {NoStop}%
\bibitem [{\citenamefont {Korenaga}(2010)}]{korenaga2010}%
  \BibitemOpen
  \bibfield  {author} {\bibinfo {author} {\bibnamefont {Korenaga}, \bibfnamefont {J.}},\ }\bibfield  {title} {\enquote {\bibinfo {title} {Scaling of plate tectonic convection with pseudoplastic rheology},}\ }\href {\doibase 10.1029/2010JB007670} {\bibfield  {journal} {\bibinfo  {journal} {Journal of Geophysical Research: Solid Earth}\ }\textbf {\bibinfo {volume} {115}} (\bibinfo {year} {2010}),\ 10.1029/2010JB007670}\BibitemShut {NoStop}%
\bibitem [{\citenamefont {Korenaga}\ and\ \citenamefont {Jordan}(2003)}]{korenaga2003}%
  \BibitemOpen
  \bibfield  {author} {\bibinfo {author} {\bibnamefont {Korenaga}, \bibfnamefont {J.}}\ and\ \bibinfo {author} {\bibnamefont {Jordan}, \bibfnamefont {T.~H.}},\ }\bibfield  {title} {\enquote {\bibinfo {title} {Physics of multiscale convection in earth's mantle: Onset of sublithospheric convection},}\ }\href {\doibase 10.1029/2002JB001760} {\bibfield  {journal} {\bibinfo  {journal} {Journal of Geophysical Research: Solid Earth}\ }\textbf {\bibinfo {volume} {108}} (\bibinfo {year} {2003}),\ 10.1029/2002JB001760}\BibitemShut {NoStop}%
\bibitem [{\citenamefont {Lam}\ \emph {et~al.}(2023)\citenamefont {Lam}, \citenamefont {Sanchez-Gonzalez}, \citenamefont {Willson}, \citenamefont {Wirnsberger}, \citenamefont {Fortunato}, \citenamefont {Alet}, \citenamefont {Ravuri}, \citenamefont {Ewalds}, \citenamefont {Eaton-Rosen}, \citenamefont {Hu}, \citenamefont {Merose}, \citenamefont {Hoyer}, \citenamefont {Holland}, \citenamefont {Vinyals}, \citenamefont {Stott}, \citenamefont {Pritzel}, \citenamefont {Mohamed},\ and\ \citenamefont {Battaglia}}]{graphcast}%
  \BibitemOpen
  \bibfield  {author} {\bibinfo {author} {\bibnamefont {Lam}, \bibfnamefont {R.}}, \bibinfo {author} {\bibnamefont {Sanchez-Gonzalez}, \bibfnamefont {A.}}, \bibinfo {author} {\bibnamefont {Willson}, \bibfnamefont {M.}}, \bibinfo {author} {\bibnamefont {Wirnsberger}, \bibfnamefont {P.}}, \bibinfo {author} {\bibnamefont {Fortunato}, \bibfnamefont {M.}}, \bibinfo {author} {\bibnamefont {Alet}, \bibfnamefont {F.}}, \bibinfo {author} {\bibnamefont {Ravuri}, \bibfnamefont {S.}}, \bibinfo {author} {\bibnamefont {Ewalds}, \bibfnamefont {T.}}, \bibinfo {author} {\bibnamefont {Eaton-Rosen}, \bibfnamefont {Z.}}, \bibinfo {author} {\bibnamefont {Hu}, \bibfnamefont {W.}}, \bibinfo {author} {\bibnamefont {Merose}, \bibfnamefont {A.}}, \bibinfo {author} {\bibnamefont {Hoyer}, \bibfnamefont {S.}}, \bibinfo {author} {\bibnamefont {Holland}, \bibfnamefont {G.}}, \bibinfo {author} {\bibnamefont {Vinyals}, \bibfnamefont {O.}}, \bibinfo {author} {\bibnamefont {Stott}, \bibfnamefont {J.}}, \bibinfo {author} {\bibnamefont
  {Pritzel}, \bibfnamefont {A.}}, \bibinfo {author} {\bibnamefont {Mohamed}, \bibfnamefont {S.}}, \ and\ \bibinfo {author} {\bibnamefont {Battaglia}, \bibfnamefont {P.}},\ }\bibfield  {title} {\enquote {\bibinfo {title} {Learning skillful medium-range global weather forecasting},}\ }\href {\doibase 10.1126/science.adi2336} {\bibfield  {journal} {\bibinfo  {journal} {Science}\ }\textbf {\bibinfo {volume} {382}},\ \bibinfo {pages} {1416--1421} (\bibinfo {year} {2023})}\BibitemShut {NoStop}%
\bibitem [{\citenamefont {Lippe}\ \emph {et~al.}(2023)\citenamefont {Lippe}, \citenamefont {Veeling}, \citenamefont {Perdikaris}, \citenamefont {Turner},\ and\ \citenamefont {Brandstetter}}]{lippe2023pderefiner}%
  \BibitemOpen
  \bibfield  {author} {\bibinfo {author} {\bibnamefont {Lippe}, \bibfnamefont {P.}}, \bibinfo {author} {\bibnamefont {Veeling}, \bibfnamefont {B.~S.}}, \bibinfo {author} {\bibnamefont {Perdikaris}, \bibfnamefont {P.}}, \bibinfo {author} {\bibnamefont {Turner}, \bibfnamefont {R.~E.}}, \ and\ \bibinfo {author} {\bibnamefont {Brandstetter}, \bibfnamefont {J.}},\ }\bibfield  {title} {\enquote {\bibinfo {title} {{PDE}-refiner: Achieving accurate long rollouts with neural {PDE} solvers},}\ }in\ \href {https://openreview.net/forum?id=Qv6468llWS} {\emph {\bibinfo {booktitle} {Thirty-seventh Conference on Neural Information Processing Systems}}}\ (\bibinfo {year} {2023})\BibitemShut {NoStop}%
\bibitem [{\citenamefont {Liu}, \citenamefont {Chu},\ and\ \citenamefont {Thuerey}(2025)}]{liu2025config}%
  \BibitemOpen
  \bibfield  {author} {\bibinfo {author} {\bibnamefont {Liu}, \bibfnamefont {Q.}}, \bibinfo {author} {\bibnamefont {Chu}, \bibfnamefont {M.}}, \ and\ \bibinfo {author} {\bibnamefont {Thuerey}, \bibfnamefont {N.}},\ }\bibfield  {title} {\enquote {\bibinfo {title} {Con{FIG}: Towards conflict-free training of physics informed neural networks},}\ }in\ \href {https://openreview.net/forum?id=APojAzJQiq} {\emph {\bibinfo {booktitle} {The Thirteenth International Conference on Learning Representations}}}\ (\bibinfo {year} {2025})\BibitemShut {NoStop}%
\bibitem [{\citenamefont {Pathak}\ \emph {et~al.}(2020)\citenamefont {Pathak}, \citenamefont {Mustafa}, \citenamefont {Kashinath}, \citenamefont {Motheau}, \citenamefont {Kurth},\ and\ \citenamefont {Day}}]{pathak2020usingmachinelearningaugment}%
  \BibitemOpen
  \bibfield  {author} {\bibinfo {author} {\bibnamefont {Pathak}, \bibfnamefont {J.}}, \bibinfo {author} {\bibnamefont {Mustafa}, \bibfnamefont {M.}}, \bibinfo {author} {\bibnamefont {Kashinath}, \bibfnamefont {K.}}, \bibinfo {author} {\bibnamefont {Motheau}, \bibfnamefont {E.}}, \bibinfo {author} {\bibnamefont {Kurth}, \bibfnamefont {T.}}, \ and\ \bibinfo {author} {\bibnamefont {Day}, \bibfnamefont {M.}},\ }\href {https://arxiv.org/abs/2010.00072} {\enquote {\bibinfo {title} {Using machine learning to augment coarse-grid computational fluid dynamics simulations},}\ } (\bibinfo {year} {2020}),\ \Eprint {http://arxiv.org/abs/2010.00072} {arXiv:2010.00072 [physics.comp-ph]} \BibitemShut {NoStop}%
\bibitem [{\citenamefont {Raissi}, \citenamefont {Perdikaris},\ and\ \citenamefont {Karniadakis}(2019)}]{pinn}%
  \BibitemOpen
  \bibfield  {author} {\bibinfo {author} {\bibnamefont {Raissi}, \bibfnamefont {M.}}, \bibinfo {author} {\bibnamefont {Perdikaris}, \bibfnamefont {P.}}, \ and\ \bibinfo {author} {\bibnamefont {Karniadakis}, \bibfnamefont {G.~E.}},\ }\bibfield  {title} {\enquote {\bibinfo {title} {Physics-informed neural networks: A deep learning framework for solving forward and inverse problems involving nonlinear partial differential equations},}\ }\href@noop {} {\bibfield  {journal} {\bibinfo  {journal} {J. Comp. Phys.}\ }\textbf {\bibinfo {volume} {378}},\ \bibinfo {pages} {686 -- 707} (\bibinfo {year} {2019})}\BibitemShut {NoStop}%
\bibitem [{\citenamefont {Ranade}\ \emph {et~al.}(2022)\citenamefont {Ranade}, \citenamefont {Hill}, \citenamefont {Ghule},\ and\ \citenamefont {Pathak}}]{ranade2022a}%
  \BibitemOpen
  \bibfield  {author} {\bibinfo {author} {\bibnamefont {Ranade}, \bibfnamefont {R.}}, \bibinfo {author} {\bibnamefont {Hill}, \bibfnamefont {D.~C.}}, \bibinfo {author} {\bibnamefont {Ghule}, \bibfnamefont {L.}}, \ and\ \bibinfo {author} {\bibnamefont {Pathak}, \bibfnamefont {J.}},\ }\bibfield  {title} {\enquote {\bibinfo {title} {A composable machine-learning approach for steady-state simulations on high-resolution grids},}\ }in\ \href {https://openreview.net/forum?id=jvFTMD5QTq} {\emph {\bibinfo {booktitle} {Advances in Neural Information Processing Systems}}},\ \bibinfo {editor} {edited by\ \bibinfo {editor} {\bibfnamefont {A.~H.}\ \bibnamefont {Oh}}, \bibinfo {editor} {\bibfnamefont {A.}~\bibnamefont {Agarwal}}, \bibinfo {editor} {\bibfnamefont {D.}~\bibnamefont {Belgrave}}, \ and\ \bibinfo {editor} {\bibfnamefont {K.}~\bibnamefont {Cho}}}\ (\bibinfo {year} {2022})\BibitemShut {NoStop}%
\bibitem [{\citenamefont {Reese}, \citenamefont {Solomatov},\ and\ \citenamefont {Moresi}(1998)}]{reese1998}%
  \BibitemOpen
  \bibfield  {author} {\bibinfo {author} {\bibnamefont {Reese}, \bibfnamefont {C.}}, \bibinfo {author} {\bibnamefont {Solomatov}, \bibfnamefont {V.}}, \ and\ \bibinfo {author} {\bibnamefont {Moresi}, \bibfnamefont {L.-N.}},\ }\bibfield  {title} {\enquote {\bibinfo {title} {{Heat transport efficiency for stagnant lid convection with dislocation viscosity: Application to Mars and Venus}},}\ }\href@noop {} {\bibfield  {journal} {\bibinfo  {journal} {Journal of Geophysical Research - Planets}\ }\textbf {\bibinfo {volume} {103}},\ \bibinfo {pages} {13643--13657} (\bibinfo {year} {1998})}\BibitemShut {NoStop}%
\bibitem [{\citenamefont {Ronneberger}, \citenamefont {Fischer},\ and\ \citenamefont {Brox}(2015)}]{ronneberger2015u}%
  \BibitemOpen
  \bibfield  {author} {\bibinfo {author} {\bibnamefont {Ronneberger}, \bibfnamefont {O.}}, \bibinfo {author} {\bibnamefont {Fischer}, \bibfnamefont {P.}}, \ and\ \bibinfo {author} {\bibnamefont {Brox}, \bibfnamefont {T.}},\ }\bibfield  {title} {\enquote {\bibinfo {title} {U-net: Convolutional networks for biomedical image segmentation},}\ }in\ \href@noop {} {\emph {\bibinfo {booktitle} {International Conference on Medical image computing and computer-assisted intervention}}}\ (\bibinfo {organization} {Springer},\ \bibinfo {year} {2015})\ pp.\ \bibinfo {pages} {234--241}\BibitemShut {NoStop}%
\bibitem [{\citenamefont {Schubert}, \citenamefont {Turcotte},\ and\ \citenamefont {Olson}(2001)}]{schubert_book}%
  \BibitemOpen
  \bibfield  {author} {\bibinfo {author} {\bibnamefont {Schubert}, \bibfnamefont {G.}}, \bibinfo {author} {\bibnamefont {Turcotte}, \bibfnamefont {D.~L.}}, \ and\ \bibinfo {author} {\bibnamefont {Olson}, \bibfnamefont {P.}},\ }\href {\doibase 10.1017/CBO9780511612879} {\emph {\bibinfo {title} {Mantle Convection in the Earth and Planets}}}\ (\bibinfo  {publisher} {Cambridge University Press},\ \bibinfo {year} {2001})\BibitemShut {NoStop}%
\bibitem [{\citenamefont {Schulz}\ \emph {et~al.}(2020)\citenamefont {Schulz}, \citenamefont {Tosi}, \citenamefont {Plesa},\ and\ \citenamefont {Breuer}}]{schulz2020}%
  \BibitemOpen
  \bibfield  {author} {\bibinfo {author} {\bibnamefont {Schulz}, \bibfnamefont {F.}}, \bibinfo {author} {\bibnamefont {Tosi}, \bibfnamefont {N.}}, \bibinfo {author} {\bibnamefont {Plesa}, \bibfnamefont {A.-C.}}, \ and\ \bibinfo {author} {\bibnamefont {Breuer}, \bibfnamefont {D.}},\ }\bibfield  {title} {\enquote {\bibinfo {title} {Stagnant-lid convection with diffusion and dislocation creep rheology: Influence of a non-evolving grain size},}\ }\href {\doibase 10.1093/gji/ggz417} {\bibfield  {journal} {\bibinfo  {journal} {Geophysical Journal International}\ }\textbf {\bibinfo {volume} {220}},\ \bibinfo {pages} {18--36} (\bibinfo {year} {2020})}\BibitemShut {NoStop}%
\bibitem [{\citenamefont {Serrano}\ \emph {et~al.}(2023)\citenamefont {Serrano}, \citenamefont {Boudec}, \citenamefont {Koupa{\"\i}}, \citenamefont {Wang}, \citenamefont {Yin}, \citenamefont {Vittaut},\ and\ \citenamefont {patrick gallinari}}]{serrano2023operator}%
  \BibitemOpen
  \bibfield  {author} {\bibinfo {author} {\bibnamefont {Serrano}, \bibfnamefont {L.}}, \bibinfo {author} {\bibnamefont {Boudec}, \bibfnamefont {L.~L.}}, \bibinfo {author} {\bibnamefont {Koupa{\"\i}}, \bibfnamefont {A.~K.}}, \bibinfo {author} {\bibnamefont {Wang}, \bibfnamefont {T.~X.}}, \bibinfo {author} {\bibnamefont {Yin}, \bibfnamefont {Y.}}, \bibinfo {author} {\bibnamefont {Vittaut}, \bibfnamefont {J.-N.}}, \ and\ \bibinfo {author} {\bibnamefont {patrick gallinari},},\ }\bibfield  {title} {\enquote {\bibinfo {title} {Operator learning with neural fields: Tackling {PDE}s on general geometries},}\ }in\ \href {https://openreview.net/forum?id=4jEjq5nhg1} {\emph {\bibinfo {booktitle} {Thirty-seventh Conference on Neural Information Processing Systems}}}\ (\bibinfo {year} {2023})\BibitemShut {NoStop}%
\bibitem [{\citenamefont {Shahnas}\ and\ \citenamefont {Pysklywec}(2020)}]{shahnas2020}%
  \BibitemOpen
  \bibfield  {author} {\bibinfo {author} {\bibnamefont {Shahnas}, \bibfnamefont {M.~H.}}\ and\ \bibinfo {author} {\bibnamefont {Pysklywec}, \bibfnamefont {R.~N.}},\ }\bibfield  {title} {\enquote {\bibinfo {title} {Toward a unified model for the thermal state of the planetary mantle: Estimations from mean field deep learning},}\ }\href {\doibase 10.1029/2019EA000881} {\bibfield  {journal} {\bibinfo  {journal} {Earth and Space Science}\ }\textbf {\bibinfo {volume} {7}} (\bibinfo {year} {2020}),\ 10.1029/2019EA000881}\BibitemShut {NoStop}%
\bibitem [{\citenamefont {Shishehbor}, \citenamefont {Hosseinmardi},\ and\ \citenamefont {Bostanabad}(2024)}]{shishehbor_parametric_2024}%
  \BibitemOpen
  \bibfield  {author} {\bibinfo {author} {\bibnamefont {Shishehbor}, \bibfnamefont {M.}}, \bibinfo {author} {\bibnamefont {Hosseinmardi}, \bibfnamefont {S.}}, \ and\ \bibinfo {author} {\bibnamefont {Bostanabad}, \bibfnamefont {R.}},\ }\bibfield  {title} {\enquote {\bibinfo {title} {Parametric encoding with attention and convolution mitigate spectral bias of neural partial differential equation solvers},}\ }\href {\doibase 10.1007/s00158-024-03834-7} {\bibfield  {journal} {\bibinfo  {journal} {Structural and Multidisciplinary Optimization}\ }\textbf {\bibinfo {volume} {67}},\ \bibinfo {pages} {128} (\bibinfo {year} {2024})}\BibitemShut {NoStop}%
\bibitem [{\citenamefont {Solomatov}(1995)}]{solomatov1995}%
  \BibitemOpen
  \bibfield  {author} {\bibinfo {author} {\bibnamefont {Solomatov}, \bibfnamefont {V.~S.}},\ }\bibfield  {title} {\enquote {\bibinfo {title} {Scaling of temperature‐ and stress‐dependent viscosity convection},}\ }\href {\doibase 10.1063/1.868624} {\bibfield  {journal} {\bibinfo  {journal} {Physics of Fluids}\ }\textbf {\bibinfo {volume} {7}},\ \bibinfo {pages} {266--274} (\bibinfo {year} {1995})}\BibitemShut {NoStop}%
\bibitem [{\citenamefont {{Stevenson}}, \citenamefont {{Spohn}},\ and\ \citenamefont {{Schubert}}(1983)}]{stevenson1983}%
  \BibitemOpen
  \bibfield  {author} {\bibinfo {author} {\bibnamefont {{Stevenson}}, \bibfnamefont {D.~J.}}, \bibinfo {author} {\bibnamefont {{Spohn}}, \bibfnamefont {T.}}, \ and\ \bibinfo {author} {\bibnamefont {{Schubert}}, \bibfnamefont {G.}},\ }\bibfield  {title} {\enquote {\bibinfo {title} {{Magnetism and thermal evolution of the terrestrial planets}},}\ }\href {\doibase 10.1016/0019-1035(83)90241-5} {\bibfield  {journal} {\bibinfo  {journal} {Icarus}\ }\textbf {\bibinfo {volume} {54}},\ \bibinfo {pages} {466--489} (\bibinfo {year} {1983})}\BibitemShut {NoStop}%
\bibitem [{\citenamefont {{Tackley}}(2008)}]{stagyy}%
  \BibitemOpen
  \bibfield  {author} {\bibinfo {author} {\bibnamefont {{Tackley}}, \bibfnamefont {P.~J.}},\ }\bibfield  {title} {\enquote {\bibinfo {title} {{Modelling compressible mantle convection with large viscosity contrasts in a three-dimensional spherical shell using the yin-yang grid}},}\ }\href {\doibase 10.1016/j.pepi.2008.08.005} {\bibfield  {journal} {\bibinfo  {journal} {Physics of the Earth and Planetary Interiors}\ }\textbf {\bibinfo {volume} {171}},\ \bibinfo {pages} {7--18} (\bibinfo {year} {2008})}\BibitemShut {NoStop}%
\bibitem [{\citenamefont {Thiriet}\ \emph {et~al.}(2019)\citenamefont {Thiriet}, \citenamefont {Breuer}, \citenamefont {Michaut},\ and\ \citenamefont {Plesa}}]{thiriet2019}%
  \BibitemOpen
  \bibfield  {author} {\bibinfo {author} {\bibnamefont {Thiriet}, \bibfnamefont {M.}}, \bibinfo {author} {\bibnamefont {Breuer}, \bibfnamefont {D.}}, \bibinfo {author} {\bibnamefont {Michaut}, \bibfnamefont {C.}}, \ and\ \bibinfo {author} {\bibnamefont {Plesa}, \bibfnamefont {A.-C.}},\ }\bibfield  {title} {\enquote {\bibinfo {title} {Scaling laws of convection for cooling planets in a stagnant lid regime},}\ }\href {\doibase 10.1016/j.pepi.2018.11.003} {\bibfield  {journal} {\bibinfo  {journal} {Physics of the Earth and Planetary Interiors}\ }\textbf {\bibinfo {volume} {286}},\ \bibinfo {pages} {138--153} (\bibinfo {year} {2019})}\BibitemShut {NoStop}%
\bibitem [{\citenamefont {Tompson}\ \emph {et~al.}(2017)\citenamefont {Tompson}, \citenamefont {Schlachter}, \citenamefont {Sprechmann},\ and\ \citenamefont {Perlin}}]{tompson2017}%
  \BibitemOpen
  \bibfield  {author} {\bibinfo {author} {\bibnamefont {Tompson}, \bibfnamefont {J.}}, \bibinfo {author} {\bibnamefont {Schlachter}, \bibfnamefont {K.}}, \bibinfo {author} {\bibnamefont {Sprechmann}, \bibfnamefont {P.}}, \ and\ \bibinfo {author} {\bibnamefont {Perlin}, \bibfnamefont {K.}},\ }\bibfield  {title} {\enquote {\bibinfo {title} {Accelerating eulerian fluid simulation with convolutional networks},}\ }in\ \href@noop {} {\emph {\bibinfo {booktitle} {Proceedings of the 34th International Conference on Machine Learning - Volume 70}}},\ \bibinfo {series and number} {ICML'17}\ (\bibinfo  {publisher} {JMLR.org},\ \bibinfo {year} {2017})\ p.\ \bibinfo {pages} {3424–3433}\BibitemShut {NoStop}%
\bibitem [{\citenamefont {Tosi}, \citenamefont {Breuer},\ and\ \citenamefont {Spohn}(2014)}]{tosi_book}%
  \BibitemOpen
  \bibfield  {author} {\bibinfo {author} {\bibnamefont {Tosi}, \bibfnamefont {N.}}, \bibinfo {author} {\bibnamefont {Breuer}, \bibfnamefont {D.}}, \ and\ \bibinfo {author} {\bibnamefont {Spohn}, \bibfnamefont {T.}},\ }\enquote {\bibinfo {title} {Evolution of planetary interiors},}\ in\ \href {\doibase 10.1016/B978-0-12-415845-0.00009-8} {\emph {\bibinfo {booktitle} {Treatise on Geophysics (Second Edition)}}},\ \bibinfo {editor} {edited by\ \bibinfo {editor} {\bibfnamefont {G.}~\bibnamefont {Schubert}}}\ (\bibinfo  {publisher} {Elsevier},\ \bibinfo {year} {2014})\ pp.\ \bibinfo {pages} {185--208}\BibitemShut {NoStop}%
\bibitem [{\citenamefont {Tosi}\ and\ \citenamefont {Padovan}(2021)}]{Tosi2021}%
  \BibitemOpen
  \bibfield  {author} {\bibinfo {author} {\bibnamefont {Tosi}, \bibfnamefont {N.}}\ and\ \bibinfo {author} {\bibnamefont {Padovan}, \bibfnamefont {S.}},\ }\bibfield  {title} {\enquote {\bibinfo {title} {{Mercury, Moon, Mars: Surface expressions of mantle convection and interior evolution on stagnant-lid bodies}},}\ }in\ \href {\doibase 10.1002/9781119528609.ch17} {\emph {\bibinfo {booktitle} {{Mantle Convection and Surface Expressions}}}},\ \bibinfo {editor} {edited by\ \bibinfo {editor} {\bibfnamefont {H.}~\bibnamefont {Marquardt}}, \bibinfo {editor} {\bibfnamefont {M.}~\bibnamefont {Ballmer}}, \bibinfo {editor} {\bibfnamefont {S.}~\bibnamefont {Cottar}}, \ and\ \bibinfo {editor} {\bibfnamefont {J.}~\bibnamefont {Konter}}}\ (\bibinfo  {publisher} {{AGU Monograph Series}},\ \bibinfo {year} {2021})\ Chap.~\bibinfo {chapter} {17}\BibitemShut {NoStop}%
\bibitem [{\citenamefont {Um}\ \emph {et~al.}(2021)\citenamefont {Um}, \citenamefont {Brand}, \citenamefont {Yun}, \citenamefont {Fei}, \citenamefont {Holl},\ and\ \citenamefont {Thuerey}}]{um2021solverinthelooplearningdifferentiablephysics}%
  \BibitemOpen
  \bibfield  {author} {\bibinfo {author} {\bibnamefont {Um}, \bibfnamefont {K.}}, \bibinfo {author} {\bibnamefont {Brand}, \bibfnamefont {R.}}, \bibinfo {author} {\bibnamefont {Yun},}, \bibinfo {author} {\bibnamefont {Fei},}, \bibinfo {author} {\bibnamefont {Holl}, \bibfnamefont {P.}}, \ and\ \bibinfo {author} {\bibnamefont {Thuerey}, \bibfnamefont {N.}},\ }\href {https://arxiv.org/abs/2007.00016} {\enquote {\bibinfo {title} {Solver-in-the-loop: Learning from differentiable physics to interact with iterative pde-solvers},}\ } (\bibinfo {year} {2021}),\ \Eprint {http://arxiv.org/abs/2007.00016} {arXiv:2007.00016 [physics.comp-ph]} \BibitemShut {NoStop}%
\bibitem [{\citenamefont {Vyas}\ \emph {et~al.}(2025)\citenamefont {Vyas}, \citenamefont {Morwani}, \citenamefont {Zhao}, \citenamefont {Shapira}, \citenamefont {Brandfonbrener}, \citenamefont {Janson},\ and\ \citenamefont {Kakade}}]{soap}%
  \BibitemOpen
  \bibfield  {author} {\bibinfo {author} {\bibnamefont {Vyas}, \bibfnamefont {N.}}, \bibinfo {author} {\bibnamefont {Morwani}, \bibfnamefont {D.}}, \bibinfo {author} {\bibnamefont {Zhao}, \bibfnamefont {R.}}, \bibinfo {author} {\bibnamefont {Shapira}, \bibfnamefont {I.}}, \bibinfo {author} {\bibnamefont {Brandfonbrener}, \bibfnamefont {D.}}, \bibinfo {author} {\bibnamefont {Janson}, \bibfnamefont {L.}}, \ and\ \bibinfo {author} {\bibnamefont {Kakade}, \bibfnamefont {S.~M.}},\ }\bibfield  {title} {\enquote {\bibinfo {title} {{SOAP}: Improving and stabilizing shampoo using adam for language modeling},}\ }in\ \href@noop {} {\emph {\bibinfo {booktitle} {The Thirteenth International Conference on Learning Representations}}}\ (\bibinfo {year} {2025})\BibitemShut {NoStop}%
\bibitem [{\citenamefont {Wandel}, \citenamefont {Schulz},\ and\ \citenamefont {Klein}(2025)}]{wandel2025metamizer}%
  \BibitemOpen
  \bibfield  {author} {\bibinfo {author} {\bibnamefont {Wandel}, \bibfnamefont {N.}}, \bibinfo {author} {\bibnamefont {Schulz}, \bibfnamefont {S.}}, \ and\ \bibinfo {author} {\bibnamefont {Klein}, \bibfnamefont {R.}},\ }\bibfield  {title} {\enquote {\bibinfo {title} {Metamizer: A versatile neural optimizer for fast and accurate physics simulations},}\ }in\ \href {https://openreview.net/forum?id=60TXv9Xif5} {\emph {\bibinfo {booktitle} {The Thirteenth International Conference on Learning Representations}}}\ (\bibinfo {year} {2025})\BibitemShut {NoStop}%
\bibitem [{\citenamefont {Wandel}, \citenamefont {Weinmann},\ and\ \citenamefont {Klein}(2020)}]{wandel2020}%
  \BibitemOpen
  \bibfield  {author} {\bibinfo {author} {\bibnamefont {Wandel}, \bibfnamefont {N.}}, \bibinfo {author} {\bibnamefont {Weinmann}, \bibfnamefont {M.}}, \ and\ \bibinfo {author} {\bibnamefont {Klein}, \bibfnamefont {R.}},\ }\bibfield  {title} {\enquote {\bibinfo {title} {Learning incompressible fluid dynamics from scratch-towards fast, differentiable fluid models that generalize},}\ }in\ \href@noop {} {\emph {\bibinfo {booktitle} {International Conference on Learning Representations}}}\ (\bibinfo {year} {2020})\BibitemShut {NoStop}%
\bibitem [{\citenamefont {Wang}\ \emph {et~al.}(2024{\natexlab{a}})\citenamefont {Wang}, \citenamefont {Ren}, \citenamefont {Zhou}, \citenamefont {Liu}, \citenamefont {Deng}, \citenamefont {Zhang}, \citenamefont {Chengze}, \citenamefont {Liu}, \citenamefont {Wang}, \citenamefont {Wang}, \citenamefont {Wen}, \citenamefont {Sun},\ and\ \citenamefont {Liu}}]{wang2024pcnet}%
  \BibitemOpen
  \bibfield  {author} {\bibinfo {author} {\bibnamefont {Wang}, \bibfnamefont {Q.}}, \bibinfo {author} {\bibnamefont {Ren}, \bibfnamefont {P.}}, \bibinfo {author} {\bibnamefont {Zhou}, \bibfnamefont {H.}}, \bibinfo {author} {\bibnamefont {Liu}, \bibfnamefont {X.-Y.}}, \bibinfo {author} {\bibnamefont {Deng}, \bibfnamefont {Z.}}, \bibinfo {author} {\bibnamefont {Zhang}, \bibfnamefont {Y.}}, \bibinfo {author} {\bibnamefont {Chengze}, \bibfnamefont {R.}}, \bibinfo {author} {\bibnamefont {Liu}, \bibfnamefont {H.}}, \bibinfo {author} {\bibnamefont {Wang}, \bibfnamefont {Z.}}, \bibinfo {author} {\bibnamefont {Wang}, \bibfnamefont {J.-X.}}, \bibinfo {author} {\bibnamefont {Wen}, \bibfnamefont {J.-R.}}, \bibinfo {author} {\bibnamefont {Sun}, \bibfnamefont {H.}}, \ and\ \bibinfo {author} {\bibnamefont {Liu}, \bibfnamefont {Y.}},\ }\bibfield  {title} {\enquote {\bibinfo {title} {P$^2$c$^2$net: {PDE}-preserved coarse correction network for efficient prediction of spatiotemporal dynamics},}\ }in\ \href@noop {} {\emph
  {\bibinfo {booktitle} {The Thirty-eighth Annual Conference on Neural Information Processing Systems}}}\ (\bibinfo {year} {2024})\BibitemShut {NoStop}%
\bibitem [{\citenamefont {Wang}\ \emph {et~al.}(2024{\natexlab{b}})\citenamefont {Wang}, \citenamefont {Yao}, \citenamefont {Guo},\ and\ \citenamefont {Gao}}]{wangmultiscalepinn}%
  \BibitemOpen
  \bibfield  {author} {\bibinfo {author} {\bibnamefont {Wang}, \bibfnamefont {Y.}}, \bibinfo {author} {\bibnamefont {Yao}, \bibfnamefont {Y.}}, \bibinfo {author} {\bibnamefont {Guo}, \bibfnamefont {J.}}, \ and\ \bibinfo {author} {\bibnamefont {Gao}, \bibfnamefont {Z.}},\ }\bibfield  {title} {\enquote {\bibinfo {title} {A practical pinn framework for multi-scale problems with multi-magnitude loss terms},}\ }\href {\doibase https://doi.org/10.1016/j.jcp.2024.113112} {\bibfield  {journal} {\bibinfo  {journal} {Journal of Computational Physics}\ }\textbf {\bibinfo {volume} {510}},\ \bibinfo {pages} {113112} (\bibinfo {year} {2024}{\natexlab{b}})}\BibitemShut {NoStop}%
\bibitem [{\citenamefont {Wassing}, \citenamefont {Langer},\ and\ \citenamefont {Bekemeyer}(2025)}]{wassing2025}%
  \BibitemOpen
  \bibfield  {author} {\bibinfo {author} {\bibnamefont {Wassing}, \bibfnamefont {S.}}, \bibinfo {author} {\bibnamefont {Langer}, \bibfnamefont {S.}}, \ and\ \bibinfo {author} {\bibnamefont {Bekemeyer}, \bibfnamefont {P.}},\ }\bibfield  {title} {\enquote {\bibinfo {title} {Adopting computational fluid dynamics concepts for physics-informed neural networks},}\ }in\ \href {\doibase 10.2514/6.2025-0269} {\emph {\bibinfo {booktitle} {AIAA SciTech 2025 Forum}}}\ (\bibinfo  {publisher} {American Institute of Aeronautics and Astronautics},\ \bibinfo {year} {2025})\BibitemShut {NoStop}%
\bibitem [{\citenamefont {Wei}\ and\ \citenamefont {Freris}(2024)}]{wei2024}%
  \BibitemOpen
  \bibfield  {author} {\bibinfo {author} {\bibnamefont {Wei}, \bibfnamefont {L.}}\ and\ \bibinfo {author} {\bibnamefont {Freris}, \bibfnamefont {N.~M.}},\ }\bibfield  {title} {\enquote {\bibinfo {title} {Multi-scale graph neural network for physics-informed fluid simulation: Multi-scale graph neural network for physics-informed fluid simulation},}\ }\href {\doibase 10.1007/s00371-024-03402-6} {\bibfield  {journal} {\bibinfo  {journal} {Vis. Comput.}\ }\textbf {\bibinfo {volume} {41}},\ \bibinfo {pages} {1171–1181} (\bibinfo {year} {2024})}\BibitemShut {NoStop}%
\bibitem [{\citenamefont {Wu}, \citenamefont {Maruyama},\ and\ \citenamefont {Leskovec}(2022)}]{wu2022learningacceleratepartialdifferential}%
  \BibitemOpen
  \bibfield  {author} {\bibinfo {author} {\bibnamefont {Wu}, \bibfnamefont {T.}}, \bibinfo {author} {\bibnamefont {Maruyama}, \bibfnamefont {T.}}, \ and\ \bibinfo {author} {\bibnamefont {Leskovec}, \bibfnamefont {J.}},\ }\href {https://arxiv.org/abs/2206.07681} {\enquote {\bibinfo {title} {Learning to accelerate partial differential equations via latent global evolution},}\ } (\bibinfo {year} {2022}),\ \Eprint {http://arxiv.org/abs/2206.07681} {arXiv:2206.07681 [cs.LG]} \BibitemShut {NoStop}%
\bibitem [{\citenamefont {Yin}\ \emph {et~al.}(2023)\citenamefont {Yin}, \citenamefont {Kirchmeyer}, \citenamefont {Franceschi}, \citenamefont {Rakotomamonjy},\ and\ \citenamefont {Gallinari}}]{yin2023continuouspdedynamicsforecasting}%
  \BibitemOpen
  \bibfield  {author} {\bibinfo {author} {\bibnamefont {Yin}, \bibfnamefont {Y.}}, \bibinfo {author} {\bibnamefont {Kirchmeyer}, \bibfnamefont {M.}}, \bibinfo {author} {\bibnamefont {Franceschi}, \bibfnamefont {J.-Y.}}, \bibinfo {author} {\bibnamefont {Rakotomamonjy}, \bibfnamefont {A.}}, \ and\ \bibinfo {author} {\bibnamefont {Gallinari}, \bibfnamefont {P.}},\ }\href {https://arxiv.org/abs/2209.14855} {\enquote {\bibinfo {title} {Continuous pde dynamics forecasting with implicit neural representations},}\ } (\bibinfo {year} {2023}),\ \Eprint {http://arxiv.org/abs/2209.14855} {arXiv:2209.14855 [cs.LG]} \BibitemShut {NoStop}%
\bibitem [{\citenamefont {Zhong}\ \emph {et~al.}(2008)\citenamefont {Zhong}, \citenamefont {McNamara}, \citenamefont {Tan}, \citenamefont {Moresi},\ and\ \citenamefont {Gurnis}}]{zhong2008}%
  \BibitemOpen
  \bibfield  {author} {\bibinfo {author} {\bibnamefont {Zhong}, \bibfnamefont {S.}}, \bibinfo {author} {\bibnamefont {McNamara}, \bibfnamefont {A.}}, \bibinfo {author} {\bibnamefont {Tan}, \bibfnamefont {E.}}, \bibinfo {author} {\bibnamefont {Moresi}, \bibfnamefont {L.}}, \ and\ \bibinfo {author} {\bibnamefont {Gurnis}, \bibfnamefont {M.}},\ }\bibfield  {title} {\enquote {\bibinfo {title} {A benchmark study on mantle convection in a 3-d spherical shell using citcoms},}\ }\href@noop {} {\bibfield  {journal} {\bibinfo  {journal} {Geochemistry, Geophysics, Geosystems}\ }\textbf {\bibinfo {volume} {9}} (\bibinfo {year} {2008})}\BibitemShut {NoStop}%
\end{thebibliography}%

\end{document}